\documentclass[reprint,amssymb, amsmath, aps, superscriptaddress, showpacs, footinbib, prb]{revtex4-1}
\usepackage{graphicx}
\usepackage{color}
\usepackage[colorlinks,breaklinks,bookmarks=true,citecolor=blue,linkcolor=red,urlcolor=blue]{hyperref}
\newcommand{\degr}{$^\circ$}

\begin{document}

\title{Frustrated magnetic planes with intricate interaction pathways in the mineral langite Cu$_4$(OH)$_6$SO$_4\cdot 2$H$_2$O}

\author{Stefan Lebernegg}
\email{st.lebernegg@gmail.com}
\affiliation{Max Planck Institute for Chemical Physics of Solids, N\"{o}thnitzer
Str. 40, 01187 Dresden, Germany}

\author{Alexander A. Tsirlin}
\affiliation{National Institute of Chemical Physics and Biophysics, 12618 Tallinn, Estonia}
\affiliation{Experimental Physics VI, Center for Electronic Correlations and Magnetism, Institute of Physics, University of Augsburg, 86135 Augsburg, Germany}

\author{Oleg Janson}
\affiliation{Institute of Solid State Physics, Technical University Vienna, Wiedner Hauptstr. 8--10/138, 1040 Vienna, Austria}

\author{G\"{u}nther J. Redhammer}
\affiliation{Department of Materials Sciences \& Physics, University Salzburg, Hellbrunnerstr. 34, 5020 Salzburg, Austria}

\author{Helge Rosner}
\email{Helge.Rosner@cpfs.mpg.de}
\affiliation{Max Planck Institute for Chemical Physics of Solids, N\"{o}thnitzer
Str. 40, 01187 Dresden, Germany}
\date{\today}

\begin{abstract}
Magnetic and crystallographic properties of the mineral langite Cu$_4$(OH)$_6$SO$_4\cdot 2$H$_2$O are \hbox{reported}. Its layered crystal structure features a peculiar spatial arrangement of spin-$\frac12$ Cu$^{2+}$ ions that arises from a combination of corner- and edge-sharing chains. Experimentally, langite orders \hbox{antiferromagnetically} at $T_N\simeq 5.7$\,K as revealed by magnetization and specific heat \hbox{measurements}. Despite this very low energy scale of the magnetic transition, langite features significantly stronger couplings on the order of 50-70\,K. Half of the Cu$^{2+}$ spins are weakly coupled and saturate around 12\,T, where the magnetization reaches 0.5\,$\mu_B$/Cu. These findings are rationalized by density-functional band-structure calculations suggesting a complex interplay of frustrated exchange \hbox{couplings} in the magnetic planes. A simplified model of coupled magnetic sublattices explains the experimental features qualitatively. To start from reliable structural data, the crystal structure of langite in the 100--280\,K temperature range has been determined by single-crystal x-ray diffraction, and the hydrogen positions were refined computationally.
\end{abstract}

\maketitle

\section{Introduction}
Low-dimensional magnets show unique diversity of crystal structures and associated spin lattices, where a plethora of quantum phenomena can be observed.\cite{FHC_CuGeO3_spin-Peierls,skyrmions,natureBose} The physics of quantum spin chains has been actively explored in Cu$^{2+}$ compounds featuring chains of corner- or edge-sharing CuO$_4$ plaquette units. The corner-sharing geometry results in uniform spin chains with a negligibly small second-neighbor coupling, as in Sr$_2$CuO$_3$,\cite{sr2cuo3,sr2cuo3_1} AgCuVO$_4$~\cite{agcuvo4} and KCuMoO$_4$(OH).\cite{nawa2015} The edge-sharing geometry is by far more common. It gives rise to competing nearest-neighbor and next-nearest-neighbor couplings, where the former ($J_1$) is typically ferromagnetic, while the latter ($J_2$) is antiferromagnetic. Such $J_1-J_2$ frustrated spin chains develop incommensurate spin correlations and helical magnetic order,\cite{masuda2004,capogna2010} although few instances of ferromagnetic intrachain spin order are known as well.\cite{sapina1990,cuas2o4} The helical spin arrangement observed in simple binary compounds CuCl$_2$~\cite{banks2009} and CuBr$_2$~\cite{cubr2} and in more complex materials like linarite PbCu(OH)$_2$SO$_4$,\cite{linarite} all being frustrated $J_1-J_2$ spin chains, may trigger electric polarization induced by the magnetic order, thus leading to multiferroic behavior.\cite{seki2008,multiferro1,*multiferro2,yasui2011} Additionally, small interactions beyond the isotropic Heisenberg model lead to an intricate magnetic phase diagram, including multipolar (three-magnon) phases, which has been studied recently.\cite{linarite13} However, the complex interplay of frustration and anisotropy needs further investigations on different systems as, e.g., LiCuVO$_4$.\cite{licuvo4_1,*licuvo4_2,*licuvo4_3,licuvo4_4}

One may naturally ask what happens when two types of spin chains, those with edge- and corner-sharing geometries, are placed next to each other within one material. Spin systems comprising several magnetic sublattices with different dimensionalities and energy scales may have very unusual low-temperature properties. When two sublattices are weakly coupled, they are, to a certain extent, independent, hence two magnetic transitions manifesting the ordering within each of the sublattices could be observed. On the other hand, the ordering within one sublattice will necessarily depend on the other sublattice, because three-dimensional (3D) long-range order typically involves interactions between the sublattices. Unusual manifestations of quantum order-from-disorder have been observed in Sr$_2$Cu$_3$O$_4$Cl$_2$~\cite{chou1997,kim1999,kastner1999,kim2001,harris2001} featuring interpenetrating square lattices with drastically different exchange couplings. In CuP$_2$O$_6$, where spins, arranged on a planar square lattice, coexist with uniform spin chains, very strong spin fluctuations are observed even below the N\'eel temperature $T_N$, and the value of $T_N$ is unusually low for a quasi-two-dimensional (2D) antiferromagnet.\cite{cup2o6} The coexistence of corner- and edge-sharing Cu$^{2+}$ chains could be even more interesting because of the different nature of spin correlations, which are expected to be antiferromagnetic collinear and helical, for the corner- and edge-sharing chains, respectively.

The respective magnetic ground state of these compounds depends very subtly on the interplay of various exchange integrals, including possible frustration and strong quantum fluctuations. In particular, in edge-sharing geometries (with Cu--O--Cu bond angles near 90$^\circ$) the leading exchange integrals and, thus, the actual magnetic model are often difficult to establish due to a pronounced dependency of the exchange on the structural details: Small changes of bond angles or minor changes of the local Cu--O environment, e.g. by attached H-atoms,\cite{clinoclase, malachite} may even swap the ground state qualitatively. Owing to the high complexity of the structure-properties relation in these compounds, the combination of experimental investigations with theoretical methods appeared to be very successful to disentangle the complicated interplay. In particular, in recent years density functional calculations have developed to a valuable tool, establishing accurate magnetic models on microscopic grounds,\cite{azurite,cubr2,cucl2} even for involved geometries like coupled edge-sharing chains or various magnetic sublattices in a single compound.

\begin{figure*}[tbp] \includegraphics[width=17.1cm]{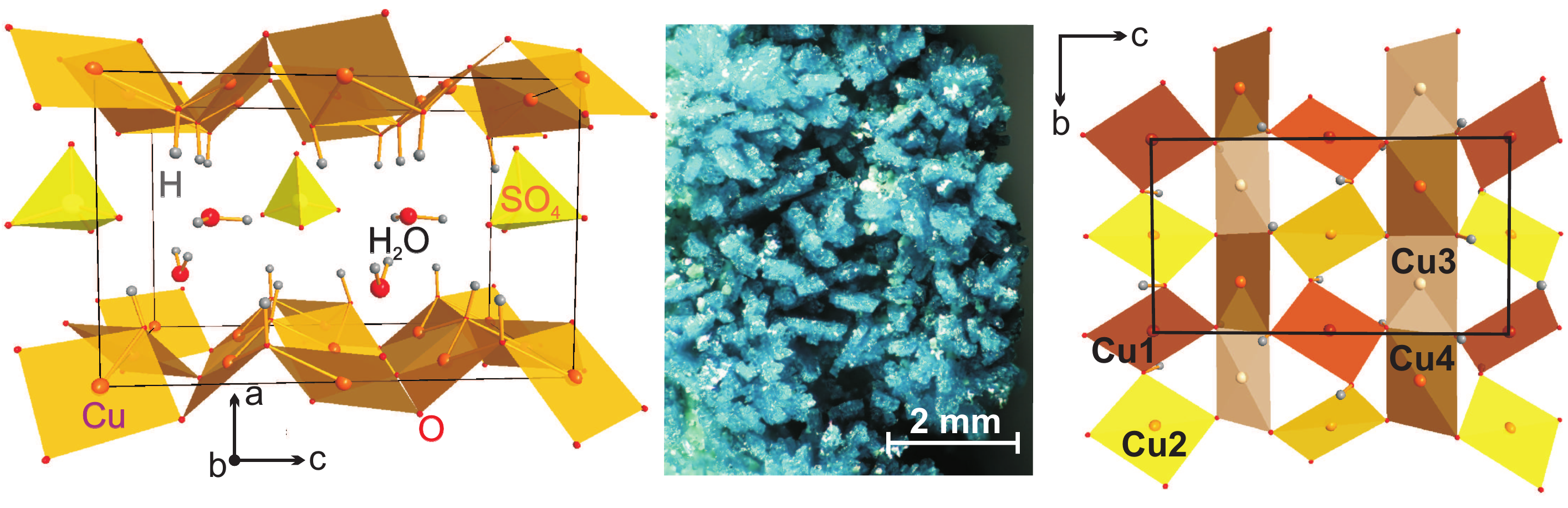}
\caption{\label{struct}(Color online) The left panel shows the crystal structure of langite, Cu$_4$(OH)$_6$SO$_4\cdot 2$H$_2$O. CuO$_4$ plaquettes are shown in orange, H are shown as grey spheres and SO$_4$ groups are shown in yellow. In the right panel, a single crystallographic layer is displayed where different colors for the four different Cu-positions are used. The picture in the center shows small light-blue langite crystals from the Podlipa and Reinera mine, Lubietova, Slovakia.}
\end{figure*}

The coexistence of the edge- and corner-sharing geometries is rather common for Cu$^{2+}$ minerals. In antlerite~\cite{antlerite2} and szenicsite,\cite{szenicsite} one edge-sharing chain is encompassed by two corner-sharing chains that together form a three-leg spin ladder. The edge- and corner-sharing chains can also form infinite layers, as in deloryite Cu$_4$(UO$_2$)(MoO$_4$)$_2$(OH)$_6$,\cite{deloryite} derriksite Cu$_4$(UO$_2$)(SeO$_3$)$_2$(OH)$_6$,\cite{derriksite} niedermayrite Cu$_4$Cd(SO$_4$)$_2$(OH)$_6$,\cite{niedermayrite} and langite Cu$_4$(OH)$_6$SO$_4\cdot 2$H$_2$O. The crystal structure of the latter mineral is shown in Fig.~\ref{struct}. Layers of alternating edge- and corner-sharing chains (Fig.~\ref{struct}, right) are well separated by SO$_4$ sulphate groups and water molecules. A somewhat similar structure without water molecules has been reported for the mineral brochantite Cu$_4$(OH)$_6$SO$_4$\cite{brochantite,brochantite2} that, however, features a much smaller interlayer separation, hence substantial interlayer couplings can be expected. In this paper, we focus on the magnetism of langite, where individual structural planes should be very weakly coupled magnetically and sufficiently pure natural samples of this mineral are available.

We also show that, in contrast to our expectations, individual structural chains in this material cannot be considered as weakly coupled magnetic chains. On the other hand, two sublattices formed by the structural chains of either type, can be distinguished in the overall very intricate spin lattice. These sublattices reveal drastically different magnetic couplings and facilitate the description of the complex low-temperature magnetism on a qualitative microscopic level.

The paper is organized as follows: Applied experimental and theoretical methods are described in Sec.~\ref{sec:methods}. The crystal structure of langite including the single-crystal data collected at low temperatures and hydrogen positions determined computationally is presented in Sec.~\ref{sec:xxstr}. Sec.~\ref{sec:exp} provides experimental results on thermodynamic properties of langite. The electronic band structure and computed exchange coupling constants are discussed in Sec.~\ref{sec:magexc}. Eventually, a detailed discussion and summary are given in Secs.~\ref{sec:disc} and~\ref{sec:summary}, respectively.

\section{Methods}
\label{sec:methods}
All experiments have been performed on a natural sample (Fig.~\ref{struct}) of langite from the Podlipa and Reinera mine, Lubietova, Slovakia. The sample quality was first thoroughly controlled by laboratory powder X-ray diffraction (XRD) (Huber G670 Guinier camera, CuK$_{\alpha\,1}$ radiation, ImagePlate detector, $2\theta\,=\,3-100^{\circ}$ angle range). 

Single-crystal X-ray diffraction between 100 and 280\,K was performed on a Bruker SMART APEX CCD-diffractometer equipped with a Cryosteam liquid nitrogen low-temperature device. A single crystal, selected on the basis of its optical properties (sharp extinctions, regular shape and homogeneity in color) was glued on top of a glass capillary (0.1 mm). Intensity data were collected with graphite-monochromatized MoK$_{\alpha}$ radiation (50\,kV, 30\,mA). The crystal-to-detector distance was 40\,mm and the detector was positioned at $-28^{\circ}$ 2$\Theta$ using an $\omega$-scan mode strategy at four different $\varphi$-positions (0$^{\circ}$, 90$^{\circ}$, 180$^{\circ}$ and 270$^{\circ}$). 630 frames with $\Delta$$\omega$\,=\,0.3$^{\circ}$ were acquired for each run. The 3D data were integrated and corrected for Lorentz polarization and background effects using the APEX2 software (Bruker -- Nonius, 2004). Structure solution (using Patterson methods) and subsequent weighted full-matrix least-square refinements on $F^2$ were done with SHELXL-97~\cite{shelx} as implemented in the program suite WinGX 1.64.\cite{wingx} 

All further experiments were performed on a powder sample since the natural crystals are very small and fragile as well as strongly intergrown, preventing us from collecting a sufficient amount of single crystallites for magnetic and specific heat measurements. Magnetization measurements were done on a Quantum Design (QD) SQUID MPMS in magnetic fields up to 5\,T and using the vibrating sample magnetometer (VSM) setup of Quantum Design PPMS up to 14\,T in the temperature range of 1.6--300\,K. Heat capacity data were acquired with the QD PPMS in fields up to 14\,T.

Electronic structure calculations within density functional theory (DFT) were performed with the full-potential local-orbital code \texttt{FPLO9.07-41}~\cite{fplo} on the 100\,K crystal structure in combination with the local density approximation (LDA),\cite{pw92} generalized gradient approximation (GGA)~\cite{pbe96} and the DFT+$U$ method.\cite{lsdau1,lsdau2} A 4$\times$4$\times$4 $k$-mesh was employed for LDA and GGA runs while super cells used for DFT+$U$ calculations were computed for about 100 $k$-points in the symmetry-irreducible part of the first Brillouin zone. We also performed auxiliary calculations using the Heyd-Scuseria-Ernzerhof (HSE06) hybrid DFT-functional\cite{hse06,*hse06a} as implemented in the Vienna Ab initio Simulation Package (\texttt{VASP5.2}) code.\cite{vasp1,*vasp2}

The hydrogen positions, which are essential for the calculation of the exchange couplings,~\cite{malachite,clinoclase} have not been determined so far~\cite{xxstr} since H is almost invisible in XRD due to its very low scattering power. Alternative experimental techniques such as neutron diffraction require large and, preferably, deuterated samples that are not available in nature. Therefore, we determined the positions of hydrogen by numerical optimization of the atomic parameters with respect to a minimization of the total energy. These calculations were performed within GGA and have proved to be highly efficient and sufficiently accurate for cuprates in recent studies.\cite{malachite,callaghanite,clinoclase}  

The exchange coupling constants $J_{ij}$ were calculated within DFT following two different strategies. One strategy involves the analysis of the half-filled LDA bands at the Fermi level allowing for the determination of leading exchange pathways by an evaluation of the electron hopping integrals $t_{ij}$. The $t_{ij}$ are computed as off-diagonal Hamiltonian matrix elements of Cu-centered Wannier functions (WFs) constructed for the half-filled bands. The spurious metallic state produced for magnetic insulators within LDA can be remedied by inserting LDA-based $t_{ij}$ into an effective Hubbard model with the effective onsite Coulomb repulsion $U_{\text{eff}}$, where in cuprates typically $U_{\text{eff}} \simeq 4.5$\,eV.\cite{malachite,cup2o6,callaghanite} In the limit of strong correlations, $t_{ij} \ll U_{\text{eff}}$, which is perfectly fulfilled in langite (see Table~\ref{tJ}), antiferromagnetic (AFM) contributions to the total exchange constants $J_{ij}$ can be estimated in second order as $J^{\text{AFM}}_{ij}=4t^2_{ij}/U_{\text{eff}}$. A more detailed description of the procedure can be found, e.g., in Refs.~\onlinecite{malachite,cux2}.

Alternatively, strong electron correlations are added on top of LDA by the LSDA+$U$ method in a mean-field way and are thus included in the self-consistent procedure. This allows for calculating total exchange constants $J_{ij}=J_{ij}^{\text{FM}}+J_{ij}^{\text{AFM}}$, which contain also the ferromagnetic (FM) contributions. A fully localized limit (FLL) approximation was used for correcting the double counting. The on-site Coulomb repulsion and Hund's exchange were set to $U_d=8.5 \pm 1$\,eV and $J_d=1$\,eV, respectively, a choice which has been successfully used for several other cuprates.\cite{azurite,malachite,callaghanite} The total exchange coupling constants $J_{ij}$ of the spin Hamiltonian 
\begin{equation} 
\hat{H}=\sum_{\left\langle ij\right\rangle}J_{ij}\hat{S_{i}}\cdot\hat{S_{j}}
\end{equation}
are calculated as differences between total energies of various collinear (broken-symmetry) spin states.\cite{callaghanite,cubr2,xiang} 

\section{Crystal structure}
\label{sec:xxstr}

Using the experimental crystal structure of langite reported in Ref.~\onlinecite{xxstr}, we first routinely performed a DFT-optimization of the atomic parameters of all atoms in the unit cell with the lattice parameters being fixed to their experimental values. Deviations up to 0.3\,\r{A} between the experimental and optimized Cu--O bond lengths prompted us to reinvestigate the crystal structure of langite with single-crystal XRD. We also performed low-temperature XRD measurements in order to probe possible temperature-induced structural changes that may be relevant to understanding the magnetism. 

Table~\ref{atomdat} compiles the results of the structural study at 100\,K, which served as input for all DFT calculations. Additional crystallographic data collected at 140, 220, 250, 280\,K are provided in the Supplementary Material.\cite{supplement} In the temperature range between 100--280\,K the unit cell volume increases by about 0.9\% with increasing $T$. The largest change in the lattice parameters was observed not along $a$ perpendicular to the structural layers (Fig.~\ref{struct}), as one might intuitively expect, but along the $c$ direction. With increasing $T$, the $c$ parameter increases by about 0.35\% arising reflecting the flattening of the layers. Changes along the $a$ and $b$ axes are similar, about 0.26\% each. The monoclinic angle remains almost constant for the investigated temperature range. 

In the presently available structural data,\cite{xxstr} hydrogen positions have been determined on a semiempirical level, only. One of the hydrogen has been placed on the sulphate group which is quite unexpected. In the related Cu-sulphate brochantite, Cu$_4$(OH)$_6$SO$_4$, H-atoms have been reliably located by neutron diffraction on a deuterated sample, and no hydrogen was found at the SO$_4$ groups but at the Cu-O layers. \cite{brochantite} More doubts on the reliability of the tentative H-positions of langite as provided in Ref.~\onlinecite{xxstr} arise from the geometry of the water molecules. While one of them shows bond lengths close to a free water molecule, the other one is strongly distorted with O--H distances of 0.919\,\r{A} and 1.032\,\r{A}, respectively, and a H--O--H angle of only 88.54°. These issues already call for a reinvestigation of the hydrogen positions in langite. Besides gaining new structural information, accurate atomic H-positions are also essential for the computation of exchange coupling constants which are very sensitive to O--H distances and the position of H with respect to the CuO$_4$ plaquette planes (Fig.~\ref{struct}).\cite{clinoclase}  
New atomic hydrogen positions are given in Table~\ref{atomdat}, which were obtained by GGA-optimization (see Sec.~\ref{sec:methods}) using various tentative positions as starting values to test the stability of the results. When only hydrogen atoms were allowed to relax, the forces on the oxygen atoms of water molecules (OW1 and OW2) turned out to be quite large, while one of the hydrogen atoms moved towards the SO$_4$ group. Though, such a situation cannot fully be excluded and may arise for a certain temperature regime due to the spatial proximity of layers and SO4 groups it appears unlikely as explained before.
In a further step, the positions of all H atoms together with those of OW1 and OW2 have been relaxed. This way, we could stabilize the anticipated langite structure by 2.6\,eV/unit cell, while the HSO$_4$ configuration became energetically highly unfavorable. A full relaxation of all atomic positions further confirms the stability of the Cu$_4$(OH)$_6$SO$_4\cdot 2$H$_2$O structure and shows no signatures of the HSO$_4$ groups. For the two different water molecules, O--H bond lengths between 0.985-0.997\,\r{A} and H--O--H angles of 103.8\degr and 109.2\degr, respectively, have been obtained, i.e. there are no asymmetrical distortions as proposed in the previous structural work.\cite{xxstr} A plot showing the hydrogen bonds and the bonding between the SO$_4$ groups, water molecules and Cu--O layers can be found in the supplementary material.\cite{supplement}
LDA band structures and density of states around the Fermi level computed for the different crystal structures, i.e. from Ref.~\onlinecite{xxstr} and our data collected at 100\,K, are provided in the supplementary material~\cite{supplement}. LDA-calculations on the crystal structures with optimized hydrogen positions and optimized positions of oxygen in the water molecules (OW) are shown as well. Band shifts between 50--100\,meV and considerable changes in the band dispersion are observed particularly between -0.6 to -0.1\,eV. Since the LDA-bands around the Fermi level crucially determine the exchange interactions, these data demonstrate how crucial hydrogen positions and accurate crystal structures are for computing a microscopic magnetic model.

Table~\ref{atomdat} summarizes atomic positions in langite, including the OW1 and OW2 positions determined both experimentally and by the GGA-optimization. The difference between the experimental and computational positions of water molecules may reflect temperature-induced structural changes, because DFT yields the crystal structure at zero temperature, whereas experimental structure determination has been performed down to 100\,K, only. However, we did not observe any sharp structural phase transitions below 100\,K in thermodynamic properties reported in Sec.~\ref{sec:exp}. It is also possible that the discrepancy between the experimental and computational positions of water molecules is intrinsic and related to marginal disorder, which is a plausible explanation, given the weak (hydrogen) bonding between the water molecules and the rest of the crystal structure. Vibration spectroscopy could provide further insight into the nature of hydrogen bonding and positions of water molecules in langite, but it lies beyond the scope of our study, which is focused on the magnetism of langite. Relevant magnetic interactions run within the Cu--O layers and should not depend on the exact positions of the out-of-plane water molecules. For the sake of consistency and given the fact that magnetic ordering in langite occurs well below 100\,K, we used the relaxed positions of OW1 and OW2 in the further microscopic analysis (Sec.~\ref{sec:magexc}).

\begin{table}[tbp]
\begin{ruledtabular}
\caption{\label{atomdat} 
Refined atomic positions (in fractions of lattice parameters) and isotropic atomic displacement parameters $U_{iso}$ (in $\times 10^{-2}$ \,\r{A}$^2$) of langite, Cu$_4$(OH)$_6$SO$_4\cdot 2$H$_2$O, collected at 100\,K. Refinement residuals are $R_1=4.64$\%, $wR_2=7.6$\%. All atoms are in the general position $2a$ of the space group $P1c1$. The lattice parameters are as follows: $a=7.1231(8)$\,\r{A}, $b=6.0305(7)$\,\r{A}, $c=11.1935(12)$\,\r{A} and $\beta=90.1479(14)^{\circ}$. OW and HW denote the O and H atoms of the H$_2$O molecules. The OS atoms belong to the SO$_4$ tetrahedra. The H-positions have been obtained by numerical optimization within GGA. For OW, experimental and optimized positions are provided.}
\begin{tabular}{c c c c c}
Atom &  $x/a$     & $y/b$        & $z/c$        & $U_{iso}$  \\ \hline
Cu1 & 0.99960(13) &  0.99762(18) &  0.49990(9)  &  0.46(2)   \\
Cu2 & 0.99260(13) &  0.49213(19) &  0.50188(9)  &  0.43(3)   \\
Cu3 & 0.00399(14) &  0.75566(18) &  0.75295(11) &  0.40(2)   \\
Cu4 & 0.00878(15) &  0.25490(17) &  0.75183(11) &  0.38(2)   \\
S   & 0.5778(4)   &  0.1854(3)   &  0.4201(3)   &  0.46(4)   \\
O1  & 0.8860(9)   &  0.0001(10)  &  0.6634(6)   &  0.41(12)  \\
O2  & 0.8882(9)   &  0.5044(10)  &  0.6652(6)   &  0.54(13)  \\
O3  & 0.1156(9)   &  0.5098(10)  &  0.8432(7)   &  0.43(13)  \\
O4  & 0.1412(8)   &  0.2452(11)  &  0.5604(6)   &  0.39(13)  \\
O5  & 0.8600(8)   &  0.7441(9)   &  0.4412(6)   &  0.45(13)  \\
O6  & 0.1256(9)   &  0.0054(10)  &  0.8398(7)   &  0.56(14)  \\
OS1 & 0.7845(8)   &  0.2311(9)   &  0.4116(5)   &  0.42(12)  \\
OS2 & 0.5388(8)   &  0.0666(9)   &  0.5344(6)   &  1.08(13)  \\
OS3 & 0.4770(8)   &  0.4015(10)  &  0.4219(6)   &  0.94(12)  \\
OS4 & 0.5165(8)   &  0.9519(10)  &  0.8163(6)   &  0.91(13)  \\
OW1 & 0.2626(9)   &  0.7398(10)  &  0.6008(6)   &  0.73(13)  \\
OW2 & 0.5178(9)   &  0.4283(11)  &  0.6955(7)   &  1.16(15)  \\
\multicolumn{4}{c}{GGA-optimization}            & \\\cline{1-4}
OW1 &  0.27422    &  0.73641     &  0.59891     & \\
OW2 &  0.52514    &  0.43403     &  0.69849     & \\
H1  &  0.27170    &  0.74411     &  0.03266     & \\ 
H2  &  0.35870    &  0.38187     &  0.12901     & \\ 
H3  &  0.72674    &  0.74906     &  0.46285     & \\ 
H4  &  0.26599    &  0.00612     &  0.33549     & \\ 
H6  &  0.74761    &  0.99558     &  0.65733     & \\ 
H7  &  0.25162    &  0.46853     &  0.36133     & \\
H1W1 &  0.51463    &  0.54313     &  0.28622     & \\
H2W1 &  0.36074    &  0.85755     &  0.57697     & \\
H1W2 &  0.51392    &  0.27508     &  0.68068     & \\
H2W2 &  0.74644    &  0.49444     &  0.67040     & \\
\end{tabular}
\end{ruledtabular}
\end{table}

The crystal structure of langite features four different Cu-positions. The basic building unit are layers formed by planar chains of edge-sharing CuO$_4$ plaquettes (type A chain) as well as buckled chains of corner-sharing CuO$_4$ plaquettes (type B chains) (Fig.~\ref{struct}), where the chains are directly linked to each other. Sulphate groups and water molecules are located between the layers. The Cu--O--Cu bridging angles, which are of crucial importance for the exchange couplings between the Cu-sites, amount to 99.49\degr/99.11\degr, 98.64\degr/97.81\degr~in the type-A chains. Between two edgesharing plaquettes in the type-A chains, the two bridging angles are different, i.e. the bridge is not symmetrical. Both angles are given separated by \"/\".  In the cornersharing type-B chains two different bridging angles occur of 101.05\degr~and 104.71\degr~, respectively (see Table~\ref{tJ}). A figure showing Cu--O bonding distances and bridging angles of the two different chain types is provided in the supplementary material.\cite{supplement} The bridging angles between the two chain types, A and B, are between 105--109\degr, i.e. the layers are strongly buckled (Fig.~\ref{struct}). According to Goodenough-Kanamori rules, one expects ferromagnetic (FM) exchange for bridging angles close to 90\degr and antiferromagnetic (AFM) exchange for larger bridging angles. The crossover is at about 95--100\degr,\cite{braden,cux2} and the exchange couplings in the edge-sharing chains of langite are difficult to guess in this transition region, even qualitatively (see Sec.~\ref{sec:magexc}), while all other couplings would naively be assumed AFM. However, such simple considerations are bound to fail for langite as will be demonstrated in Secs.~\ref{sec:magexc} and~\ref{sec:disc} below.

\section{Thermodynamic properties}
\label{sec:exp}
All measurements presented in this section were performed on powder from the same specimen as the one used for the single-crystal XRD. The powder quality has been diligently checked by powder XRD, revealing almost pure langite.\cite{supplement}

\begin{figure}[tb] \includegraphics[width=8.6cm]{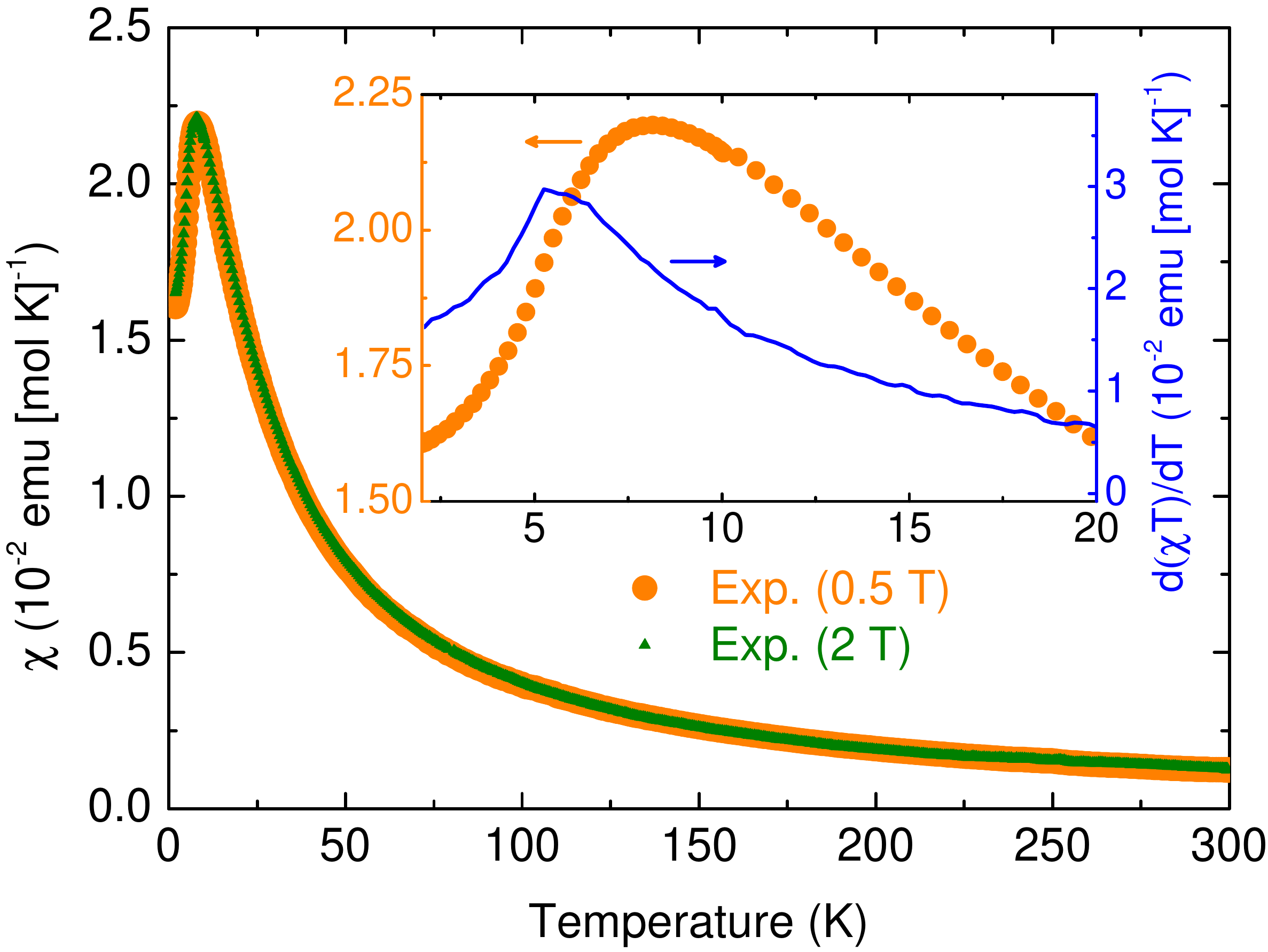}
\caption{\label{sus}(Color online) The experimental susceptibility data $\chi(T)$ of langite, Cu$_4$(OH)$_6$SO$_4\cdot 2$H$_2$O, collected at magnetic fields of 0.5 and 2.0\,T on a powder sample. $d(\chi T)/dT$ is shown in the inset as blue line.}
\end{figure}

The temperature-dependent magnetic susceptibility $\chi(T)$ measured in magnetic fields of 0.5 and 2\,T is shown in Fig.~\ref{sus}, where the two curves are almost identical. A Curie-Weiss fit $\chi(T)=C/(T+\theta)$ of the high-temperature regime (220--290\,K) of the 2\,T data yields $\theta=18.2$\,K and $C=0.481$\,emu K (mol Cu)$^{-1}$. From the constant $C$, we obtain an effective magnetic moment of 1.96 $\mu_B$/Cu which is larger than the spin only value of 1.73 and implies the $g$-factor of 2.26 lying still in the expected range for Cu$^{2+}$ compounds.\cite{malachite,diaboleite,brochantite} The positive $\theta$ indicates predominant antiferromagnetic couplings, which are, however, quite weak.
In the low-temperature regime, $\chi(T)$ features a rather sharp peak at 7.5\,K. This peak is somewhat asymmetric and thus different from the susceptibility maxima in conventional low-dimensional antiferromagnets, where short-range magnetic order is formed well above the N\'eel temperature $T_N$.\cite{yogi2015,ahmed2015} While no indications of a magnetic transition are seen in the raw susceptibility data, Fisher's heat capacity $d(\chi T)/dT$ reveals a kink around 5.5\,K that can be paralleled to the anomaly in the specific heat and ascribed to the magnetic ordering transition. The absence of a Curie tail at the lowest $T$, typically arising from paramagnetic spin-1/2 impurities (see e.g. Ref.~\onlinecite{malachite}), demonstrates the high quality of our natural sample. 

\begin{figure}[tb] \includegraphics[width=8.6cm]{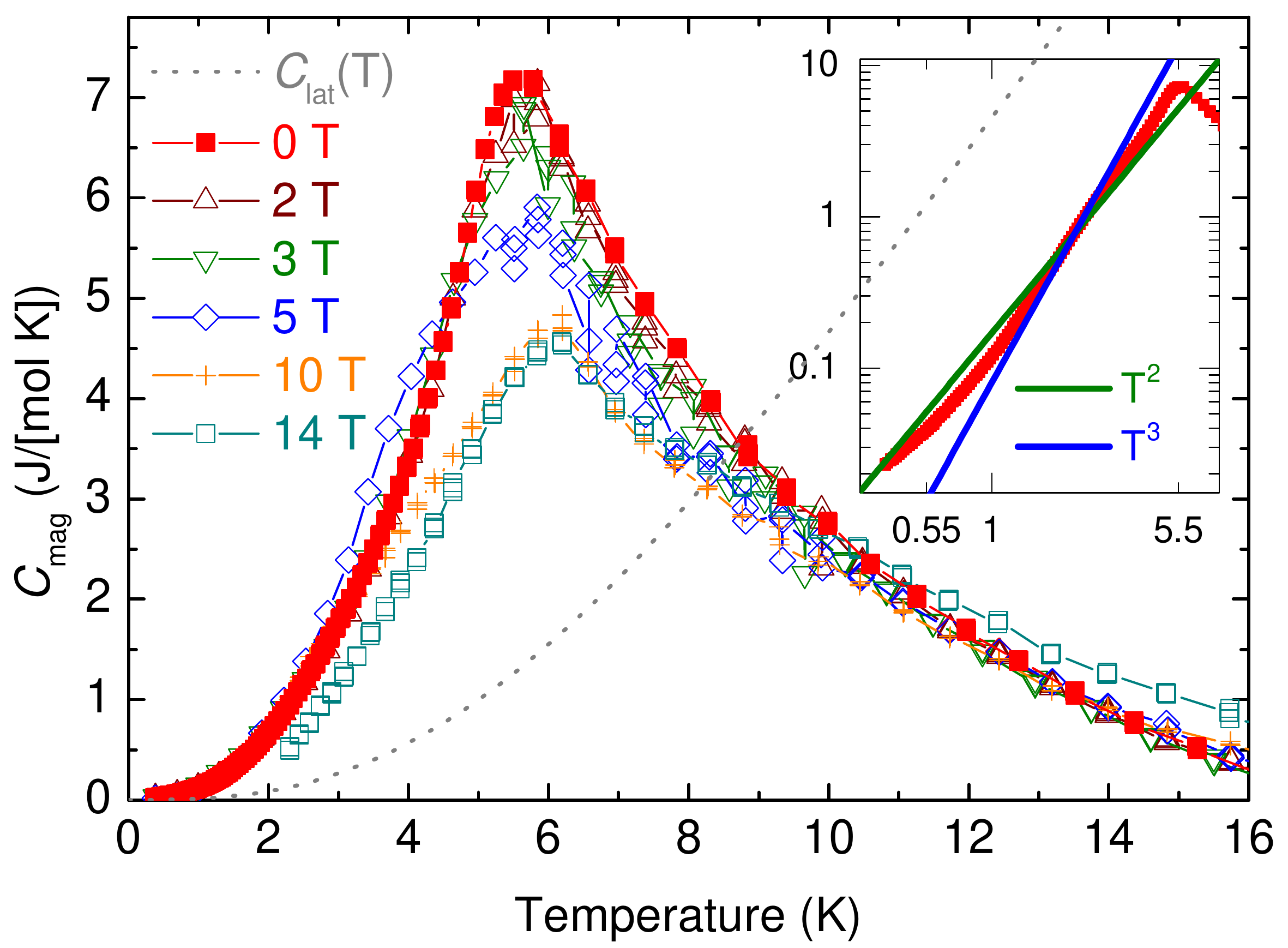}
\caption{\label{cp}(Color online) The magnetic contribution to the specific heat, $C_{\text{mag}}(T)$, of langite, Cu$_4$(OH)$_6$SO$_4\cdot 2$H$_2$O, measured in magnetic fields between 0--14\,T on a powder sample. The inset shows a double logarithmic plot of the magnetic specific heat $C_{mag}(T)$ of langite collected in zero magnetic field. The green and blue lines show $T^2$ and $T^3$ fits, respectively. $T^2$ and $T^3$ would be the anticipated behaviors of a two-dimensional and three-dimensional antiferromagnet, respectively. The grey dotted curve shows the fitted lattice background $C_{lat}(T)$.}
\end{figure}

The 5.7\,K anomaly in the specific heat generally resembles a $\lambda$-type anomaly, which is expected at a second-order phase transition. The broadening of this anomaly may be driven by effects of magnetic anisotropy. Magnetic nature of the 5.7\,K transition is corroborated by its field dependence. Despite the relatively low value of $T_N$, the transition is well visible up to at least 14\,T, and the transition temperature changes only slightly in the applied magnetic field. At higher temperatures, the lattice contribution to the specific heat dominates. Below the transition temperature $T_N$, $C_{mag}(T)$ decreases, but it does not follow the simple $T^2$ behavior of a 2D antiferromagnet which might be anticipated for a layered system like langite. However, it is also not following a $T^3$ dependence expected for a conventional 3D antiferromagnet (see Fig.\ref{cp}).

The magnetic contribution $C_{\text{mag}}$ was obtained by subtracting the lattice contribution $C_{\text{lat}}$ from the measured $C_p$ data, where $C_{\text{lat}}(T)$ was approximated by fitting a polynomial~\cite{supplement} $C_{\text{lat}}(T)=\sum \limits_{n=3}^{n=7} c_nT^n$, proposed by Johnston \textit{et al.},\cite{johnston2000} to the $C_p(T)$ data in the temperature range of 20--39\,K. The same polynomial was used to extract the magnetic contributions from specific heat data measured in the various magnetic fields. In zero magnetic field, the magnetic entropy $S_{\text{mag}}$, released within the magnetic transition, was estimated to about 6.8\,J/(mol K) by integrating $C_{\text{mag}}/T$. Thus, only about 30\% of the expected $S_{\text{mag}}=4R\ln 2$ for a spin-1/2 system is released within the transition anomaly and right above $T_N$, while the rest is spread towards higher temperatures, which is typical for low-dimensional antiferromagnets\cite{ahmed2015} and corroborates that $T_N$ is somewhat lower than the energy scale of the exchange couplings given by, e.g., $\theta\simeq 18$\,K. A similar value for $S_{\text{mag}}$ has been reported for the related mineral brochantite (see also Sec.\ref{sec:disc}) releasing 7.9\,J/(mol K), which is about 34\% of the total magnetic entropy, in the vicinity of the magnetic transition.\cite{brochantite} 

\begin{figure}[tb] \includegraphics[width=8.6cm]{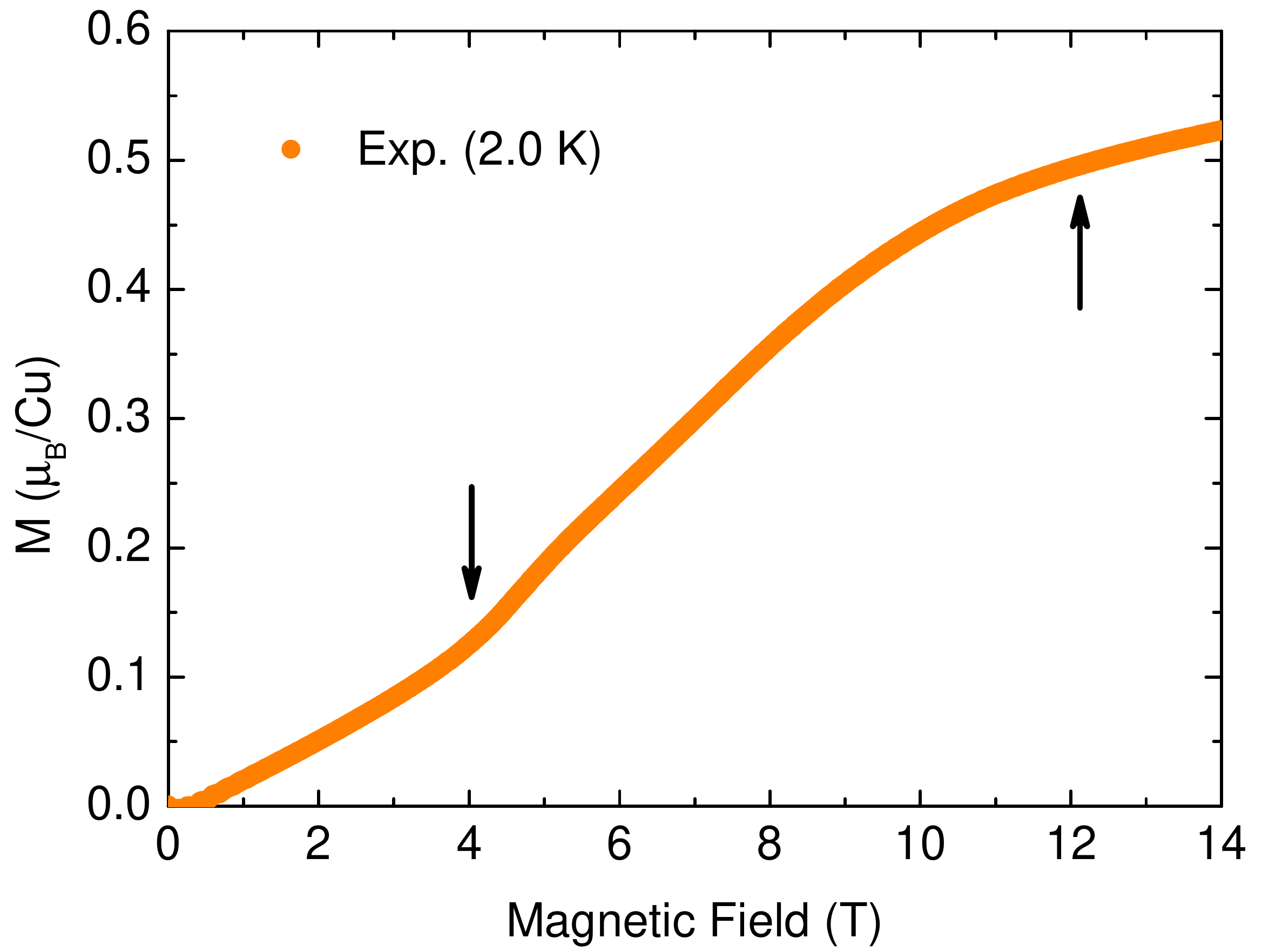}
\caption{\label{mH}(Color online) Field-dependent magnetization data M(H) of langite, Cu$_4$(OH)$_6$SO$_4\cdot 2$H$_2$O, up to 14\,T collected at a temperature of 2.0\,K on a powder sample. The arrows indicate the kink at about 4\,T and half saturation which is reached at about 12\,T assuming a $g$-factor of 2.0.}
\end{figure}

Field-dependent magnetization $M(H)$ (Fig.~\ref{mH}) measured in fields up to 14\,T features a kink around 4\,T, reaches half-saturation around 12\,T, and keeps increasing up to at least 14\,T. The kink at 4\,T is reminiscent of a spin-flop transition that, however, happens at a much higher field than in other Cu$^{2+}$ oxides.\cite{[{For example:}][{}]arango2011} Above 2\,K, the features of the magnetization curve are smeared out, so we were not able to map them as a function of temperature and construct a comprehensive $T-H$ phase diagram.

\section{Electronic structure and magnetic exchange couplings}
\label{sec:magexc}

In this section, we derive a microscopic magnetic model that could be used to understand the complex behavior of langite. 
Microscopic models based on empirical considerations are prone to error because superexchange in Cu$^{2+}$ compounds depends on tiny structural details and cannot be fully captured by empirical rules. Moreover, the presence of four distinct Cu sites in the crystal structure implies that interactions with similar Cu--Cu distances and superexchange pathways are not related by symmetry and may be unequal. Therefore, an empirical approach for deriving a microscopic magnetic model is bound to fail for langite. Accordingly, we employ numerical electronic structure DFT calculations allowing for a direct computation of individual exchange couplings $J_{ij}$. In combination with numerical simulations of the thermodynamical properties, such calculations often provided consistent description of the macroscopic magnetic behavior based on microscopic considerations.\cite{cup2o6,cubr2,cucl2,a2cup2o7}

With suitably chosen correlation parameters, such as the Coulomb repulsion $U_d$ in LSDA+$U$, one expects that DFT results are accurate within 10\% for most insulating spin-1/2 materials and the respective interaction pathways. However, the error bars increase for very weak couplings and for those couplings, where special nature of the superexchange pathway renders ferro- and antiferromagnetic contributions comparable in size. Further information on the computational procedure and the accuracy of computed exchange couplings can be found in Refs.~\onlinecite{bs2,hth75,bs1,cubr2,azurite,cdvo3}.

\begin{figure}[tb] \includegraphics[width=8.6cm]{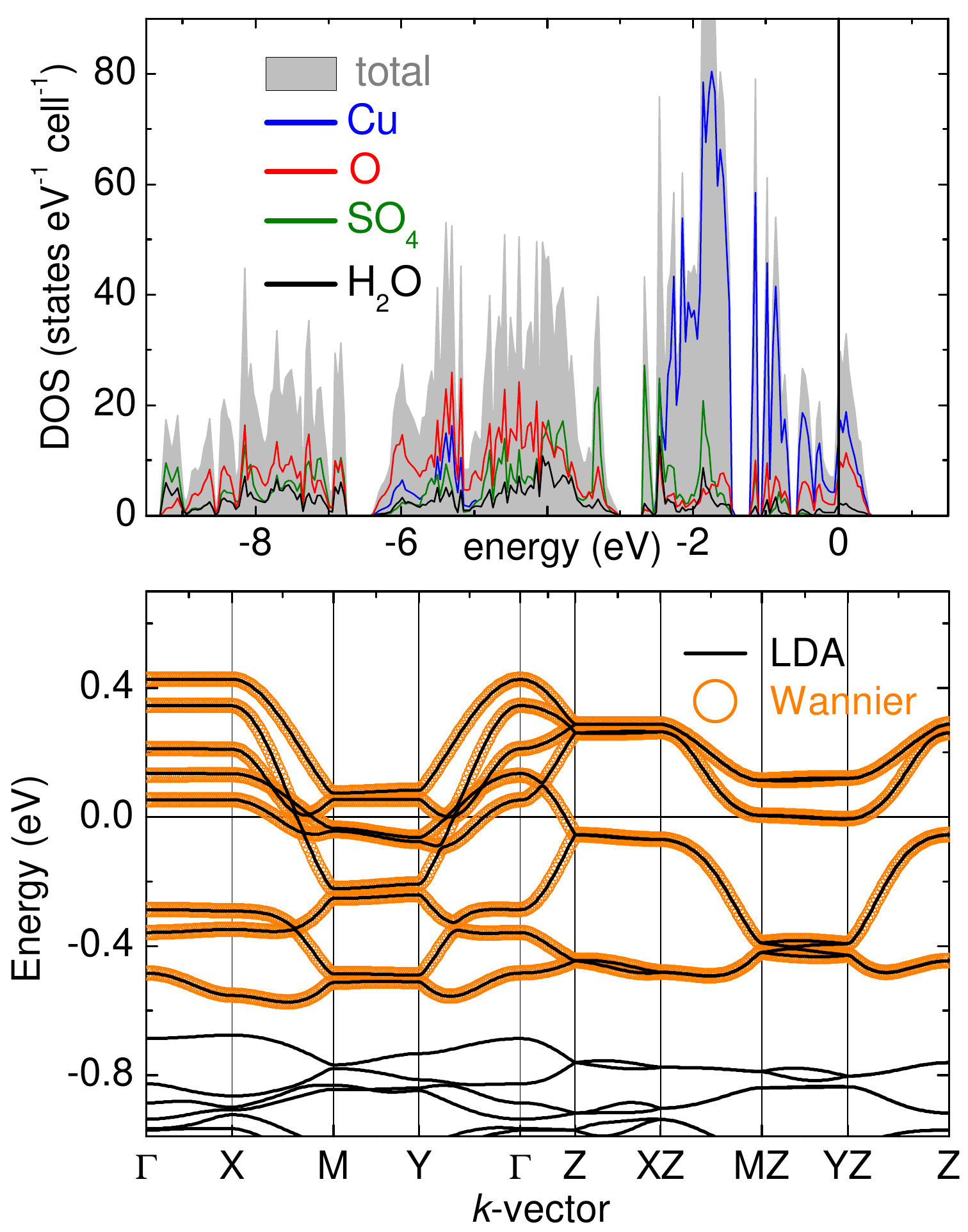}
\caption{\label{lda}(Color online) The top panel shows the total and partial density of states (DOS) from LDA calculations. In the lower panel the eight half-filled LDA bands around the Fermi level are shown. "Wannier" denotes bands calculated with Cu-centered Wannier functions.}
\end{figure}

\begin{figure*}[tbp] \includegraphics[width=17.1cm]{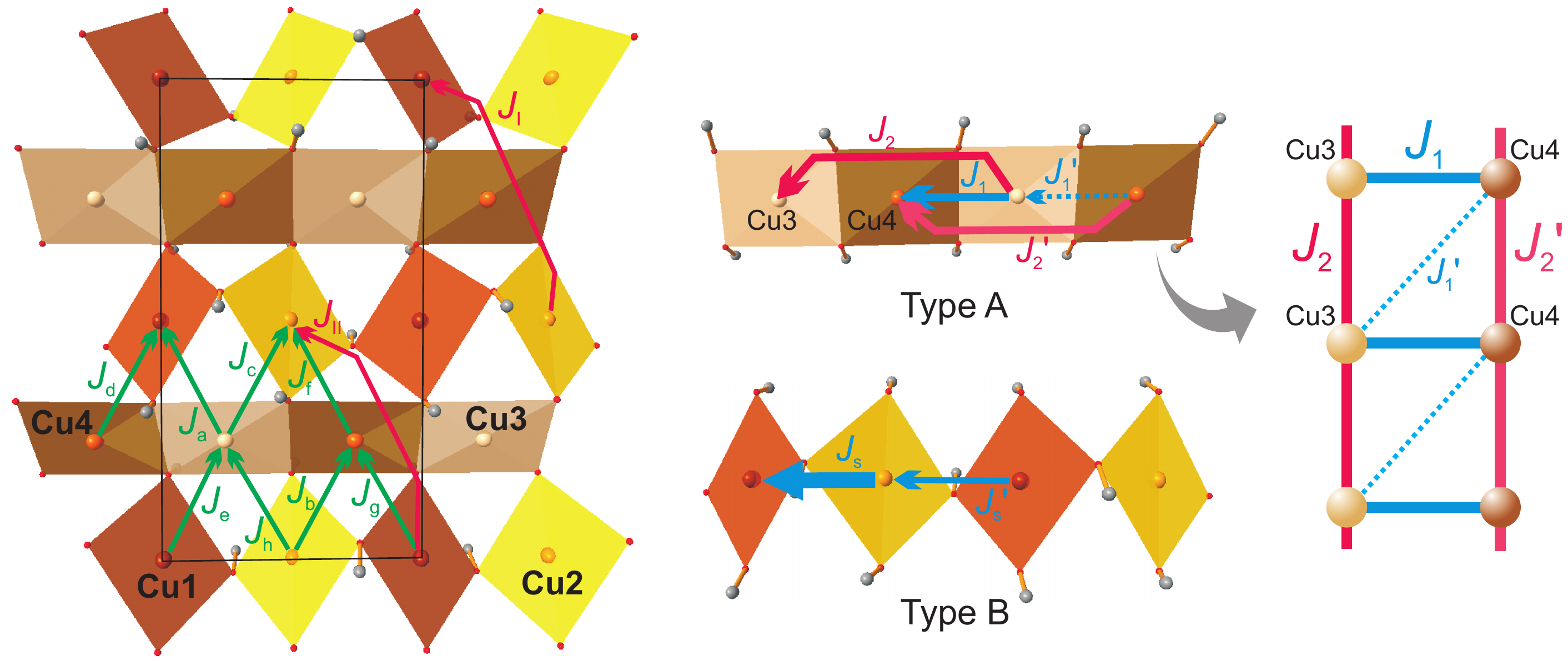}
\caption{\label{hop}(Color online) The left panel shows the intralayer exchange pathways. The central panel shows the two structural chain types and the respective intrachain exchange pathways. The exchange interactions of Cu-spins on the type-B chains may be described in terms of alternating ferromagnetic spin chains which are coupled with each other by $J_{II}$. This represents the first magnetic sublattice (SL1). A two-leg spin ladder drawn from the intrachain exchange couplings of the Cu-spins on the type-A chains is shown in the right panel. These antiferromagnetic spin ladders form the second magnetic sublattice (SL2). A list of all exchange couplings, their strengths and structural characteristics are provided in Table~\ref{tJ}.}
\end{figure*}

As a first step, LDA calculations were performed, yielding a broad valence band complex of about 10\,eV (Fig.~\ref{lda}), which is typical for cuprates.\cite{clinoclase,malachite,callaghanite} Low-energy magnetic excitations should be largely determined by the band complex of eight half-filled bands around the Fermi level, between $-0.5$ and 0.45\,eV. The eight bands arise from the eight Cu$^{2+}$-ions per unit cell and their corresponding eight half-filled $3d$-orbitals. Local coordinate systems on the eight Cu-sites (with the local $z$-axis chosen perpendicular to the CuO$_4$-planes and the local $x$-axis oriented parallel to a Cu--O bond) allow analyzing the orbital character of the half-filled bands. They are essentially of Cu($3d_{x^2-y^2}$) and O($2p_x$,$2p_y$) character while admixtures from H$_2$O and particularly SO$_4$ groups are small. Accordingly, the latter molecules do not play a direct role for the exchange couplings in langite. This set of eight bands is now projected onto Cu-centered Wannier functions (WFs) to evaluate the hopping parameters $t_{ij}$. Owing to the four different Cu-sites, many different exchange pathways are effective in langite. Table~\ref{tJ} lists all $|t_{ij}|>20$\,meV as well as the corresponding $J^{\text{AFM}}_{ij}$.  The largest interlayer hopping $t'$ is only about $-5$\,meV rendering the spin lattice of langite nearly two-dimensional. The positions of the respective exchange pathways in the crystal structure are shown in Fig.~\ref{hop}.

\begin{table}[tbp]
\begin{ruledtabular}
\caption{\label{tJ} 
The transfer integrals $t_{ij}$ (meV) and the AFM contributions to the exchange constants $J^{\text{AFM}}_{ij}=4t_{ij}^2/U_{\text{eff}}$ (K) where $U_{\text{eff}}$\,=\,4.5\,eV. d(Cu--Cu) and Cu--O--Cu denote Cu--Cu distances (\r{A}) and Cu--O--Cu angles (deg), respectively. The $J_{ij}$ (K) given in the last column are calculated with the LSDA+$U$ method and $U_d=8.5\pm1$\,eV and $J_d=1$\,eV. The different groups of exchange couplings are sorted with respect to Cu--Cu distances. For a detailed explanation of the bridging angles see Sec.~\ref{sec:xxstr} as well as the supplementary material.\cite{supplement}}
\begin{tabular}{c c c c c c}
          &  d(Cu--Cu)  &   Cu--O--Cu    & $t_{ij}$ & $J^{\text{AFM}}_{ij}$  & $J_{ij}$        \\ \hline
\multicolumn{6}{c}{type A chains (edge-sharing)} \\
$J_1'$    & 3.011       &	97.81/98.64	 & 	$-144$     &   213                   &    $9\pm15$     \\
$J_1$     & 3.020       &	99.10/99.49	 &  $-155$     &   247                   &   $38\pm20$     \\
$J_2'$    & 6.030       &	             &    61       &    39                   &   $36\pm6$      \\
$J_2$     & 6.030       &	             &    59       &    36                   &   $35\pm6$      \\
\multicolumn{6}{c}{type B chains (corner-sharing)} \\
$J_s$     & 2.983       &	101.05	     &    40       &    16                   &  $-74\pm10$     \\
$J_s'$    & 3.049       &	104.706	     &    62       &    40                   &  $-23\pm3$      \\
$J_{s2}$  & 6.030       &	             &    31       &    10                   &   $<1$          \\
$J_{s2}'$ & 6.030       &	             &    30       &     9                   &   $<1$          \\
\multicolumn{6}{c}{nearest-neighbor interchain couplings} \\
$J_a$     & 3.139       &	103.55       &    85       &    74                   &  $-19\pm1$      \\
$J_b$     & 3.144       &	105.08       &  $-109$     &   123                   &    $5\pm5$      \\
$J_c$     & 3.163       &	106.64       &    91       &    85                   &   $-6\pm2$      \\
$J_d$     & 3.168       &	105.23       &   109       &   121                   &    $8\pm5$      \\
$J_e$     & 3.186       &	106.65       & 	 $-91$     &    85                   &    $0\pm4$      \\
$J_f$     & 3.190       &	106.61       &    94       &    91                   &  $-12\pm2$      \\
$J_g$     & 3.219       &	106.10       &   $-92$     &    87                   &  $-13\pm1$      \\
$J_h$     & 3.229       &	109.23       & 	$-107$     &   119                   &   $31\pm7$      \\
\multicolumn{6}{c}{next-nearest-neighbor interchain couplings} \\
$J_I$     & 6.309       &	             &   $-51$     &    27                   &   $10\pm10$     \\
$J_{II}$  & 6.407	      &              &   $-61$     &    38                   &   $51\pm20$     \\
\multicolumn{6}{c}{interlayer coupling} \\
$J'$      & 7.874       &              &    $-5$     &     0.2                 &                 \\
\end{tabular}
\end{ruledtabular}
\end{table}

Full exchange constants $J_{ij}=J_{ij}^{\text{FM}}+J_{ij}^{\text{AFM}}$ computed with the LSDA+$U$ method are provided in the last column of Table~\ref{tJ}, where error bars show the effect of changing the Coulomb repulsion parameter $U_d$ by $\pm1$\,eV. This parameter affects absolute values of exchange couplings, while their ratios typically change by few percent only. However, for weak couplings error bars can exceed absolute values, and thus the ratios are strongly affected as well. 

The two nearest-neighbor (NN) couplings of the Cu-spins on the type-B chains, $J_s$ and $J_s'$, are both FM, with $J_s'$ being much smaller than $J_s$. This results in ferromagnetic spin chains with alternating exchange couplings. These spin chains interact antiferromagnetically via $J_{II}$ and represent the first magnetic sublattice (SL1). The difference between $J_s$ and $J_s'$ can be traced back to the relevant Cu--O--Cu angles (Table~\ref{tJ}). The smaller angle for $J_s$ leads to a stronger FM interaction (see also Sec.~\ref{sec:discussion}).

The NN coupling $J_1$ of the Cu-spins on the edge-sharing type-A chains is AFM and about four times stronger than the weak coupling $J_1'$. The AFM next-nearest-neighbor (NNN) couplings $J_2$ and $J_2'$ are of the same strength as $J_1$. The exchange interactions of the spins on the type-A chains may, thus, be described in terms of magnetic two-leg ladders (Fig.~\ref{hop}), where $J_2$ and $J_2'$ build the legs, while $J_1$ forms the rungs, and $J_1'$ is a frustrating diagonal interaction. This represents the second magnetic sublattice (SL2). The strengths of $J_1$ and $J_1'$ can be again traced back to the Cu--O--Cu angles. The smaller bridging angles render $J_1'$ weaker than $J_1$. However, these \emph{antiferromagnetic} couplings are observed for the bridging angles below 100$^{\circ}$, while the \emph{ferromagnetic} couplings $J_s$ and $J_s'$ occur for the bridging angles above 100$^{\circ}$. This instructive situation highlights limitations of the Goodenough-Kanamori-Anderson rules and the importance of the mutual arrangement of the CuO$_4$ plaquettes, which share edges ($J_1$ and $J_1'$) or corners ($J_s$ and $J_s'$), respectively.

Multiple couplings between the two sublattices are mostly weak. The strongest inter-sublattice interaction $J_h$ is comparable in size to $J_1$, $J_2$, and $J_2'$. However, it is less abundant than the intra-sublattice couplings, and on average one finds that only half of $J_h$ contributes to the effective molecular field on the SL2. Therefore, in a first approximation one can consider langite as antiferromagnet built of two sublattices, where the sublattice SL2 is 1D, the sublattice SL1 is 2D, and the inter-sublattice couplings are weaker than the leading couplings within each of the sublattices. The interlayer coupling $J'$, which is about 0.2\,K, may be responsible for the long-range magnetic order observed in langite because three-dimensional order requires the coupling between the layers (Fig.~\ref{cp}). We refrained from estimating $J'$ using LSDA+$U$, though, because such small couplings are hard to calculate reliably.\cite{callaghanite}

The exchange couplings given in Table~\ref{tJ} allow estimating a "local Curie-Weiss temperature" $\theta_k$ for each Cu-site $i$ according to $\theta_k=S(S+1)/3 \cdot \sum\limits_i(z_i \cdot J_i)$ where $S$ is the electron spin and $z_i$ shows how often a certain coupling $J_i$ occurs on a given site. $\theta_k$ is thereby a measure for the local coupling strengths on the Cu-site $k$. Accordingly, we get for the four Cu-sites (Fig.~\ref{hop}): $\theta_1=-15$\,K, $\theta_2=-3$\,K, $\theta_3=31$\,K, $\theta_4=27$\,K.
The overall Curie-Weiss temperature may be approximated by averaging over all sites yielding $\theta=10$\,K which is in reasonable agreement with the experimental value of $18.2$\,K (see Sec.~\ref{sec:exp}) regarding the intricate microscopic magnetic model and the large number of exchange couplings. The striking difference between $\theta_1$ and $\theta_2$ arises from the nearest-neighbor interchain couplings, where the ferromagnetic $J_a$ operates on Cu-site 1 and the antiferromagnetic $J_h$ is effective on Cu-site 2. It is worth noting that the single experimental parameter $\theta$ is usually sufficient for verifying the choice of $U_d$ in LSDA+$U$, because the uncertainty in absolute values of the computed exchange couplings is much higher than in their ratios, and thus only the absolute scale of computed exchange couplings should be cross-checked experimentally. On the other hand, experimental evaluation of individual exchange couplings in langite may be an arduous task, given the overall complexity of the spin lattice. In Sec.~\ref{sec:disc}, we further discuss ramifications of our microscopic magnetic model and its relevant macroscopic features that can be tracked experimentally.

We also performed calculations using a HSE06 hybrid DFT-functional,\cite{hse06,*hse06a} as implemented in \texttt{VASP5.2}.\cite{vasp1,*vasp2} These calculations were feasible for short-range couplings only. The long-range couplings would require big supercells that cannot be treated with the computationally expensive HSE06 method with the required accuracy. 

In contrast to LSDA+$U$, the hybrid-functional approach does not include the effect of local Coulomb repulsion explicitly, thus leading to less accurate estimates of individual exchange couplings and the overestimate of ferromagnetic terms.\cite{callaghanite} On the other hand, this method is free from adjustable parameters and does not involve the ambiguous choice of the Coulomb repulsion $U_d$. The HSE06 results can be found in the supplementary material.\cite{supplement} They are generally similar to those from LSDA+$U$ (Table~\ref{tJ}) and confirm main features of the langite spin lattice: i) the FM nature of $J_s$ and $J_s'$; ii) the AFM nature of $J_1$ and $J_1'$, and iii) the $|J_s|>|J_s'|$ and $J_1'<J_1$ trends discussed above. The inter-sublattice couplings are mostly FM in HSE06 because of the general tendency of hybrid functionals to overestimate ferromagnetic contributions to the exchange.

Disregarding the weak coupling $J_1'$, we find that individual magnetic sublattices of langite are non-frustrated because none of the leading couplings $J_s$, $J_s'$, and $J_{II}$ for the type-A chains and $J_1$, $J_2$, and $J_2'$ for the type-B chains compete to each other. The two leading couplings between the sublattices, FM $J_a$ and AFM $J_h$, are not frustrated either, because they are compatible with the AFM order between the FM type-B chains, as imposed by $J_{II}$. However, other inter-sublattice couplings frustrate the spin lattice and render it very complex. Given the large number of non-equivalent exchange couplings and their frustrated nature, we restrict ourselves to a qualitative discussion of the magnetic behavior in Sec.~\ref{sec:disc} below.

\section{Discussion}
\label{sec:disc}
\label{sec:discussion}
Magnetic properties of the natural Cu$^{2+}$-mineral langite are peculiar, yet complicated. Its crystal structure consists of layers formed by directly connected and alternating ordered edge- and corner-sharing chains of CuO$_4$ plaquettes. Such structural motives can be found in several compounds (see Sec.~\ref{sec:xxstr}) that have same topology of the magnetic layer, but slightly different Cu--O--Cu angles and, thus, potentially different exchange scenarios. Magnetic properties of only one of these materials, brochantite, have been reported. Therefore, it is still an open question which magnetic properties arise when edge- and corner-sharing Cu$^{2+}$ chains are joined into layers, and how these properties are affected by structural details. 

Our study shows that such layers cannot be viewed as a simple combination of weakly coupled magnetic chains. Unanticipated interchain couplings, such as $J_{II}$ that features an unusually long superexchange pathway, render the spin lattice much more complex. Remarkably, though, we can still split this lattice into two sublattices composed of type-A and type-B chains, respectively. The difference between these sublattices and individual chains pertains to the fact that all type-B chains form a single 2D sublattice SL1, whereas sublattice SL2 comprises weakly coupled type-A chains and thus remains effectively 1D. We should also emphasize that the inter-sublattice couplings are clearly non-negligible. A quantitative description of langite will, therefore, require the consideration of the full spin lattice that is partially frustrated. This problem must be tackled with advanced simulation techniques and lies beyond the scope of our present study, where we restrict ourselves to the qualitative analysis and demonstrate that the model of two different sublattices can be used to rationalize main features of the experimental data.

First, the abrupt increase of the magnetization in low fields and the fact that half-saturation is reached already at $10-12$\,T is consistent with the presence of sublattice SL1, which is largely ferromagnetic. This sublattice should saturate as soon as magnetic field overcomes the effect of the AFM coupling $J_{II}$. There is only one $J_{II}$ coupling per Cu site, so the half-saturation should be reached at $H_{s1}=k_BJ_{II}/(g\mu_B)\simeq 38$\,T, which is much higher than 12\,T observed experimentally. The origin of this discrepancy is not entirely clear. The presence of the second magnetic sublattice SL2 (comprising the spins on the type-A chains) and the frustrated interactions between the sublattices may overcome the effect of $J_{II}$ and facilitate the saturation of the SL1 (consisting of spins on the type-B chain) already in low field, although a detailed investigation of this behavior requires numerical simulations for the full spin lattice of langite, which are not feasible, as we explained above. The SL2 is antiferromagnetic and its saturation is expected at $H_{s2}=k_B/(g\mu_B)(J_1+J_2+J_2')\simeq 81$\,T which would be interesting to probe experimentally. We thus expect that above 14\,T the magnetization of langite increases much slower than in low fields, and that full saturation is reached around 80\,T reflecting the presence of sizable AFM couplings in this system.

Magnetic susceptibility of langite lacks a broad maximum that would be expected in a quasi-2D antiferromagnet. This observation is also consistent with the presence of the mostly FM sublattice SL1 (Fig.~\ref{hop}) which lacks any susceptibility maximum down to $T_N$ and, thus, masks the susceptibility maximum related to the AFM sublattice SL2. A similar behavior has been observed in CuP$_2$O$_6$,\cite{cup2o6} where none of the sublattices is ferromagnetic, but very weak couplings in one of the sublattices render half of the spins paramagnetic down to low temperatures, and no susceptibility maximum is observed down to $T_N$. It is worth noting that the asymmetric maximum in the susceptibility of langite around 8\,K cannot be taken as a typical signature of short-range order in a quasi-2D system, because this maximum is observed at temperatures well below the Curie-Weiss temperature $\theta\simeq 18$\,K, whereas in a 2D system, e.g., in a square-lattice antiferromagnet, $T^{\max}\simeq \theta$ is expected.

The temperature of the antiferromagnetic ordering in langite, $T_N\simeq 5.7$\,K, is quite low compared to leading exchange couplings $|J_s|$, $J_1$, $J_{II}$, $J_2$, and $J_2'$ that are at least $35-40$\,K each. We tentatively find $T_N/\bar J<0.2$, which is very low for a quasi-2D antiferromagnet,\cite{sim_tn_j} although a correct definition of an effective intralayer coupling $\bar J$ may be difficult in this case given the very complex nature of the spin lattice. 

A frustration ratio of $\theta/T_N\simeq 3$ is less impressive, but one has to acknowledge that the macroscopic $\theta$ is a sum of FM and AFM couplings (Table~\ref{tJ}) and thus underestimates the overall energy scale of exchange couplings in langite.

The magnetic ground state of langite may be peculiar. In sublattice SL1, one expects FM order along $b$ and AFM order along $a$ and $c$, arising from the interchain interaction $J_{II}$ and the weak interlayer coupling $J'$. The sublattice SL2 is a two-leg spin ladder and, when taken on its own, features a spin-singlet ground state without long-range magnetic order. Although interchain couplings and the couplings to SL1 will trigger the formation of ordered moments even in SL2, these moments are expected to be much smaller than in SL1. This difference in the ordered magnetic moments is one of the fingerprints of the two-sublattice model and can be probed experimentally by nuclear magnetic resonance or neutron scattering.

The drastic difference between the ordered magnetic moments on different Cu-sites has been previously seen in other Cu$^{2+}$ minerals. In antlerite Cu$_3$(OH)$_4$SO$_4$,\cite{antlerite2,antlerite3} two side chains of the B-type encompass the central chain of the A-type that together form a ribbon, which is sometimes considered as a three-leg spin ladder. Neutron scattering revealed ordered magnetic moment of 0.88\,$\mu_B$ on the terminal (type-B) chains and zero magnetic moment on the central (type-A) chains.\cite{antlerite2} A similar type of magnetic order is expected in langite, where spins in the type-B chains will form long-range magnetic order, whereas spins in the type-A chains should develop a gapped ground state with zero ordered moment, as typical for two-leg spin ladders. This unusual, partial magnetic order may be reflected in magnetic excitations and macroscopic properties such as specific heat below $T_N$. Indeed, the specific heat of langite clearly deviates from the standard $T^2$ or $T^3$ behaviors and remains an interesting problem for future investigation.

Another Cu$^{2+}$ mineral, brochantite Cu$_4$(OH)$_6$SO$_4$, is remarkably different from both langite and antlerite. From the chemistry perspective, it is a dehydrated version of langite featuring same type of magnetic layers. However, details of their geometry are somewhat different because water molecules are missing, and the separation between the layers is about twice shorter than in langite. Neutron diffraction reports very small magnetic moments within the corner-sharing type-B chains (0.22\,$\mu_B$) and much larger ordered moments within the type-A chains (0.74\,$\mu_B$).\cite{brochantite} This is very different from the ground state of antlerite (and, presumably, of langite) and may indicate a different exchange topology. Indeed, the Curie-Weiss temperature of brochantite ($\theta\simeq 90$\,K)\cite{brochantite} is much higher than 18\,K and 4\,K in langite and antlerite,\cite{hara2011} respectively. Moreover, brochantite features a broad susceptibility maximum around 60\,K, far above $T_N$, while neither langite nor antlerite show such broad maxima. These features suggest that magnetic interactions in brochantite are predominantly AFM, whereas langite and antlerite reveal a subtle interplay of FM and AFM exchange couplings. Further microscopic insight into these differences is clearly needed and requires a systematic computational study of the aforementioned Cu$^{2+}$ minerals. 

Naively, the AFM nature of brochantite can be ascribed to the larger Cu--O--Cu angles in the range $105.6-122.5^{\circ}$, while the Cu--O--Cu angles in langite are, generally, smaller ($103.5-109.2^{\circ}$). However, this simple analysis in the spirit of Goodenough-Kanamori-Anderson rules may often be misleading. 

We have mentioned that the values of the Cu--O--Cu bridging angles account for $|J_s|>|Js'|$ and $J_1>J_1'$, but they do not explain why $J_s$ and $J_s'$ are FM, while $J_1$ and $J_1'$ with the smaller bridging angles are AFM. Other effects are obviously important in this case. In particular, hydrogen atoms bonding to the bridging oxygen have strong influence on the superexchange.\cite{clinoclase} Hydrogen atoms located out of the CuO$_4$ planes, favor FM exchange. This is definitely relevant for $J_s$ and $J_s'$ in langite where the O--H bond on the bridging oxygen and the CuO$_4$ plaquettes enclose angles up to 60\degr. For the Cu-mineral clinoclase, Cu$_3$(AsO$_4$)(OH)$_3$ it was recently demonstrated~\cite{clinoclase} that such a large out-of-plane angle can drive the exchange coupling even from a strongly AFM to the FM coupling regime. 
Quite similar results for the intrachain physics of the type-B chains have been reported in recent studies on antlerite, Cu$_3$(OH)$_4$SO$_4$.\cite{antlerite2,antlerite3} The crystal structure of this compound features triple chains consisting of a central type-A chain and type-B chains bonded to it on each side. For the Cu-spins on the type-B chains an alternating FM coupling has been reported from neutron experiments with an antiparallel order between the chains.\cite{antlerite2,antlerite3}. For brochantite, Cu$_4$(OH)$_6$SO$_4$, featuring structural layers similar to those in langite, neutron data also revealed a FM coupling of the Cu-spins within the type-B chains.\cite{brochantite} Eventually, in a joint experimental and theoretical study~\cite{szenicsite} on the rare Cu-mineral szenicsite, Cu$_3$MoO$_4$(OH)$_4$, an alternating FM coupling on the type-B chains has been reported. This compound features triple chains similar to those in antlerite.

\section{Summary}
\label{sec:summary}
In summary, structural and magnetic properties of the Cu$^{2+}$ mineral langite have been investigated in a joint experimental and theoretical study. Crystal structure of langite was refined in the $100-280$\,K temperature range using single-crystal XRD, and the H-positions were subsequently determined for the 100\,K structure using theoretical DFT-approach. The crystal structure consists of two types of directly connected Cu-chains, edge- and corner-sharing, which form layers separated from each other by about 7.5\,\r{A}. These layers are a common structural motive in cuprate minerals, but their relevant magnetic interactions and resulting magnetic properties have been only scarcely investigated. Along with the fact that both chain-types taken on their own have revealed fascinating magnetic properties, it intrigued us what kind of physics may arise from their combination into layers. 

Our density-functional calculations show that such layers can not be viewed as a stack of weakly coupled magnetic chains. While different chains form different magnetic sublattices, interactions between the chains are non-negligible, and even the two-sublattice model describes the magnetic behavior only qualitatively. It does, however, capture the crucial feature that the sublattice B is predominantly ferromagnetic and prone to the formation of the long-range order, whereas the sublattice A is entirely antiferromagnetic and gapped because of its two-leg-ladder geometry. Therefore, we expect a peculiar magnetic ground state with drastically different ordered moments in the two sublattices. This ground state can be paralleled to that of antlerite, where the ``idle-spin'' behavior (no detectable ordered moment) on type-A chains has been observed.

Experimentally, langite undergoes long-range magnetic ordering, but at the N\'eel temperature $T_N\simeq 5.7$\,K that is well below the Curie-Weiss temperature $\theta\simeq 18$\,K. An effective ``frustration ratio'' $\theta/T_N\simeq 3$ demonstrates that the magnetic order in langite is impeded. However, the Curie-Weiss temperature is a sum of ferromagnetic and antiferromagnetic couplings and thus underestimates the energy scale of magnetic exchange. Taking computed $J$'s from Table~\ref{tJ}, one finds that the N\'eel temperature of langite is remarkably low for a quasi-2D antiferromagnet. This reduced value of $T_N$ is a signature of strong quantum fluctuations that have three concurrent origins: i) spin-$\frac12$ nature of Cu$^{2+}$ and magnetic low-dimensionality; ii) in-plane frustration; iii) proximity of sublattice A to the spin-singlet state without long-range magnetic order.

Altogether, langite is a frustrated quasi-2D antiferromagnet that reveals interesting manifestations of quantum magnetism and a peculiar two-sublattice structure of the spin lattice. Its ground state is of particular interest for future studies, given the anticipated difference between the ordering processes in two magnetic sublattices. Specific heat of langite measured in the ordered state does not follow conventional $T^2$ or $T^3$ behavior, thus providing first evidence for the unconventional nature of the magnetic ground state and calling for further investigation of this interesting material.

\acknowledgments
We are grateful to S.-L. Drechsler and S. Nishimoto for valuable discussions. We acknowledge the experimental support by Yurii Prots, Horst Borrmann (laboratory XRD) and Christoph Klausnitzer (low-temperature specific heat). We would also like to thank Prof. Klaus Thalheim and the Senckenberg
Naturhistorische Sammlung Dresden for providing the langite sample from Slovakia (inventory number 18940 Sy). OJ was partly supported by the European Research Council under the European Unions Seventh Framework Program FP7/ERC through grant agreement n.306447. AT was supported by the Mobilitas program of the ESF (grant number MTT77) and by the Alexander von Humboldt Foundation and Federal Ministry for Education and Research, Germany, through the Sofja Kovalevskaya Award.


%

\clearpage

\begin{table*}[h!]
\begin{tabular}{c}
\huge{\texttt{Supporting Material}} \\
\end{tabular}
\end{table*}

\begin{widetext}

\begin{table}[h]
\begin{ruledtabular}
\caption{\label{atomdat1} 
Obtained lattice parameters of Langite, details of the refinement procedure and settings for the single-crystal XRD at various temperatures.}
\begin{tabular}{l c c c c c}
T    (K)          &   280 	           &    250 	         &  220 	         &  140 	         & 100               \\ \hline
\multicolumn{2}{l}{\textbf{Crystallographic data}}   & 	     &             		 &    	             &                   \\ 
S.G.              &   $Pc$	           &    $Pc$	         &  $Pc$	         &  $Pc$	         &  $Pc$             \\
$a$ (\r{A})       &   7.1412(8)	       &    7.1370(8)	     &  7.1349(8)	     &  7.1255(8)	     &  7.1231(8)        \\
$b$ (\r{A})       &   6.0468(7)	       &    6.0427(7)	     &  6.0400(7)	     &  6.0312(7)	     &  6.0305(7)        \\
$c$ (\r{A})       &   11.2328(12)      &	11.2238(12)	     &  11.2204(12)	     &  11.1996(12)	     &  11.1935(12)      \\
$\beta$ (deg)     &   90.0716	       &    90.0743(14)	     &  90.0892(14)	     &  90.1279(14)	     &  90.1479814)      \\
Volume(\r{A}$^3$) &   485.048	       &    484.05(9)	     &  483.54(9)	     &  481.31(9)	     &  480.83(9)        \\
Z	              &   2	               &    2	             &  2	             &  2	             &  2                \\
Dx (mg/m$^3$)	  &   3.274	           &    3.281	         &  3.285	         &  3.300	         &  3.303            \\
Absorption (cm$^{-1}$) &   8.924	   &    8.942	         &  8.951	         &  8.993	         &  9.002            \\
Radiation	      &   Mo(K$_{\alpha}$) &	Mo(K$_{\alpha}$) &  Mo(K$_{\alpha}$) &  Mo(K$_{\alpha}$) &  Mo(K$_{\alpha}$) \\
\\
\multicolumn{2}{l}{\textbf{Data collection}}  &           &                &    	        &                    \\		
Theta range (deg)	    &   2.85--29.00	 &  2.85--29.01	  &  2.86--28.87   &  2.86--28.97	&  2.86--28.97       \\
Reflection collected	&   5811	     &  5773	      &  5764	       &  5726	        &  5657              \\
Independent reflections	&   2366	     &  2358	      &  2359	       &  2355         	&  2334              \\
Completeness of data	&   99.8	     &  99.8	      &  99.8	       &  99.8	        &  99.8              \\
$R_{int}$ (\%)          &   1.94	     &  1.95	      &  2.62	       &  4.1	        &  5.2               \\
$R_1$ ($I>2\sigma(I)$)	&   3.28	     &  3.46	      &  3.85	       &  5.2	        &  4.64              \\
$wR_2$                  &   6.33	     &  6.84	      &  7.39	       &  8.01	        &  6.77              \\
Largest diff. peak/hole (e$\times$\r{A}$^3$)	&0.53/-0.75	     &  0.99/-0.86	  &  1.12/-0.63    &  0.97/-0.90	&  1.16/-0.88        \\
\end{tabular}
\end{ruledtabular}
\end{table}

\begin{table}[h]
\begin{ruledtabular}
\caption{\label{atomdat1} 
Refined atomic positions (in fractions of lattice parameters) and isotropic atomic displacement parameters $U_{iso}$ ($\times 10^{-2}$ \,\r{A}$^2$) of Langite collected at 140\,K. All atoms are in the general position $2a$ of the space group $P1c1$. OW denotes O of the H$_2$O molecules. OS atoms belong to the SO$_4$ tetrahedra.}
\begin{tabular}{c c c c c}
Atom &  $x/a$     & $y/b$        & $z/c$        & $U_{iso}$  \\ \hline
Cu1 & 0.99970(15) &  0.9975(2)   &  0.49967(11) &  0.61(3)   \\
Cu2 & 0.99299(15) &  0.4927(2)   &  0.50197(10) &  0.55(3)   \\
Cu3 & 1.00363(16) &  0.7559(2)   &  0.75290(12) &  0.53(2)   \\
Cu4 & 1.00899(16) &  0.25469(18) &  0.75168(12) &  0.47(2)   \\
S   & 0.5775(5)   &  0.1846(4)   &  0.4204(3)   &  0.82(4)   \\
O1  & 0.8870(10)  &  0.9993(11)  &  0.6637(7)   &  0.68(14)  \\
O2  & 0.8874(10)  &  0.5052(11)  &  0.6661(7)   &  0.50(14)  \\
O3  & 1.1155(10)  &  0.5094(11)  &  0.8439(7)   &  0.59(15)  \\
O4  & 1.1402(9)   &  0.2462(12)  &  0.5597(7)   &  0.48(14)  \\
O5  & 0.8583(9)   &  0.7436(10)  &  0.4428(6)   &  0.63(14)  \\
O6  & 1.1253(9)   &  1.0059(11)  &  0.8407(7)   &  0.40(15)  \\
OS1 & 0.7837(9)   &  0.2316(10)  &  0.4119(6)   &  0.61(13)  \\
OS2 & 0.5374(9)   &  0.0648(11)  &  0.5333(7)   &  1.38(15)  \\
OS3 & 1.4760(10)  &  0.4033(11)  &  0.4229(7)   &  1.27(15)  \\
OS4 & 1.5146(10)  &  0.9525(11)  &  0.8181(7)   &  1.55(15)  \\
OW1 & 1.2595(10)  &  0.7411(11)  &  0.6016(8)   &  0.94(15)  \\
OW2 & 0.5157(11)  &  0.4272(13)  &  0.6982(8)   &  1.60(17)  \\
\end{tabular}
\end{ruledtabular}
\end{table}

\begin{table}[h]
\begin{ruledtabular}
\caption{\label{atomdat2} 
Refined atomic positions (in fractions of lattice parameters) and isotropic atomic displacement parameters $U_{iso}$ ($\times 10^{-2}$ \,\r{A}$^2$) of Langite collected at 220\,K. All atoms are in the general position $2a$ of the space group $P1c1$. OW denotes O and H atoms of the H$_2$O molecules. OS atoms belong to the SO$_4$ tetrahedra.}
\begin{tabular}{c c c c c}
Atom &  $x/a$     & $y/b$        & $z/c$        & $U_{iso}$  \\ \hline
Cu1 & 0.99948(9)  &  0.99780(11) &  0.49992(6) &  0.926(18)  \\
Cu2 & 0.99321(9)  &  0.49226(12) &  0.50203(6) &  0.897(19)  \\
Cu3 & 1.00292(10) &  0.75525(12) &  0.75311(7) &  0.892(15)  \\
Cu4 & 1.00889(10) &  0.25436(11) &  0.75165(7) &  0.830(17)  \\
S   & 0.5780(3)   &  0.1835(2)   &  0.4194(2)  &  1.15(3)    \\
O1  & 0.8890(7)   &  0.9981(6)   &  0.6620(5)  &  0.97(9)    \\
O2  & 0.8896(7)   &  0.5053(6)   &  0.6646(4)  &  0.88(9)    \\
O3  & 1.1182(7)   &  0.5106(6)   &  0.8425(5)  &  0.92(10)   \\
O4  & 1.1406(6)   &  0.2465(7)   &  0.5599(4)  &  0.85(10)   \\
O5  & 0.8597(7)   &  0.7453(6)   &  0.4432(4)  &  0.98(10)   \\
O6  & 1.1249(6)   &  1.0072(6)   &  0.8388(4)  &  0.69(10)   \\
OS1 & 0.7825(7)   &  0.2318(6)   &  0.4116(4)  &  1.13(9)    \\
OS2 & 0.5400(6)   &  0.0616(8)   &  0.5320(5)  &  2.49(11)   \\
OS3 & 1.4768(6)   &  0.3973(7)   &  0.4229(4)  &  1.96(10)   \\
OS4 & 1.5172(7)   &  0.9495(8)   &  0.8164(5)  &  2.44(11)   \\
OW1 & 1.2635(7)   &  0.7403(7)   &  0.6001(5)  &  1.82(10)   \\
OW2 & 0.5190(8)   &  0.4213(8)   &  0.6976(5)  &  2.61(11)   \\
\end{tabular}
\end{ruledtabular}
\end{table}

\begin{table}[h]
\begin{ruledtabular}
\caption{\label{atomdat3} 
Refined atomic positions (in fractions of lattice parameters) and isotropic atomic displacement parameters $U_{iso}$ ($\times 10^{-2}$ \,\r{A}$^2$) of Langite collected at 250\,K. All atoms are in the general position $2a$ of the space group $P1c1$. OW denotes O and H atoms of the H$_2$O molecules. OS atoms belong to the SO$_4$ tetrahedra.}
\begin{tabular}{c c c c c}
Atom &  $x/a$     & $y/b$        & $z/c$        & $U_{iso}$  \\ \hline
Cu1 & 0.99955(8) &  0.99755(10) &  0.49986(6) &  0.944(16)  \\
Cu2 & 0.99336(8) &  0.49193(10) &  0.50201(5) &  0.878(17)  \\
Cu3 & 1.00281(9) &  0.75523(11) &  0.75317(6) &  0.897(14)  \\
Cu4 & 1.00877(9) &  0.25437(9)  &  0.75180(6) &  0.805(15)  \\
S   & 0.5780(3)  &  0.1835(2)   &  0.4193(2)  &  1.24(2)    \\
O1  & 0.8878(6)  &  0.9990(5)   &  0.6620(4)  &  0.78(8)    \\
O2  & 0.8886(6)  &  0.5053(5)   &  0.6637(4)  &  0.85(8)    \\
O3  & 1.1165(6)  &  0.5096(6)   &  0.8416(4)  &  0.96(9)    \\
O4  & 1.1403(6)  &  0.2455(6)   &  0.5596(4)  &  0.87(8)    \\
O5  & 0.8608(6)  &  0.7446(5)   &  0.4427(4)  &  0.83(8)    \\
O6  & 1.1250(6)  &  1.0060(5)   &  0.8381(4)  &  0.78(9)    \\
OS1 & 0.7835(6)  &  0.2322(5)   &  0.4113(4)  &  1.00(8)    \\
OS2 & 0.5399(6)  &  0.0620(7)   &  0.5316(4)  &  2.51(9)    \\
OS3 & 1.4780(6)  &  0.3970(6)   &  0.4231(4)  &  2.08(9)    \\
OS4 & 1.5165(6)  &  0.9503(7)   &  0.8158(4)  &  2.49(10)   \\
OW1 & 1.2626(6)  &  0.7402(6)   &  0.6006(4)  &  1.72(9)    \\
OW2 & 0.5193(7)  &  0.4239(8)   &  0.6992(5)  &  2.65(10)   \\
\end{tabular}
\end{ruledtabular}
\end{table}

\begin{table}[h]
\begin{ruledtabular}
\caption{\label{atomdat4} 
Refined atomic positions (in fractions of lattice parameters) and isotropic atomic displacement parameters $U_{iso}$ ($\times 10^{-2}$ \,\r{A}$^2$) of Langite collected at 280\,K. All atoms are in the general position $2a$ of the space group $P1c1$. OW denotes O and H atoms of the H$_2$O molecules. OS atoms belong to the SO$_4$ tetrahedra.}
\begin{tabular}{c c c c c}
Atom &  $x/a$     & $y/b$        & $z/c$        & $U_{iso}$  \\ \hline
Cu1 & 0.9997(1) &  0.9976(1) &  0.4999(0) &  0.983(14)  \\
Cu2 & 0.9932(1) &  0.4919(1) &  0.5019(0) &  0.968(15)  \\
Cu3 & 1.0028(1) &  0.7552(1) &  0.7531(1) &  0.938(12)  \\
Cu4 & 1.0090(1) &  0.2541(1) &  0.7517(1) &  0.874(14)  \\
S   & 0.5778(3) &  0.1832(2) &  0.4193(2) &  1.35(2)  \\
O1  & 0.8878(5) &  0.9988(4) &  0.6631(4) &  0.99(7)  \\
O2  & 0.8882(5) &  0.5036(5) &  0.6648(4) &  0.92(7)  \\
O3  & 1.1168(5) &  0.5096(5) &  0.8427(4) &  0.94(8)  \\
O4  & 1.1407(5) &  0.2452(5) &  0.5600(4) &  1.01(8)  \\
O5  & 0.8608(5) &  0.7446(4) &  0.4433(3) &  0.91(7)  \\
O6  & 1.1243(5) &  1.0060(5) &  0.8392(4) &  0.99(8)  \\
OS1 & 0.7832(5) &  0.2308(5) &  0.4109(3) &  1.17(7)  \\
OS2 & 0.5399(5) &  0.0618(6) &  0.5312(4) &  2.63(8)  \\
OS3 & 1.4770(5) &  0.3950(6) &  0.4232(4) &  2.36(8)  \\
OS4 & 1.5171(6) &  0.9498(6) &  0.8156(4) &  2.64(9)  \\
OW1 & 1.2620(6) &  0.7399(6) &  0.6010(4) &  2.03(9)  \\
OW2 & 0.5177(6) &  0.4177(7) &  0.7000(5) &  3.12(10)  \\
\end{tabular}
\end{ruledtabular}
\end{table}

\begin{table}[h]
\begin{ruledtabular}
\caption{\label{atomdat4} 
Selected bond lengths (\r{A}) for the crystal structures of Langite collected at different temperatures.}
\begin{tabular}{c c c c c c c}
T (K)&      &  280 	    &  250   	&  220     	&  140 	    &  100       \\   
Cu1	&  O4	&  1.927(3)	&  1.925(4)	&  1.923(4)	&  1.924(7)	&  1.924(7)  \\
	&  O6	&  1.931(3)	&  1.931(4)	&  1.930(4)	&  1.940(6)	&  1.938(6)  \\
	&  O1	&  2.000(4)	&  1.987(5)	&  1.982(6)	&  2.007(8)	&  2.003(7)  \\
	&  O7	&  2.013(4)	&  2.024(5)	&  2.017(5)	&  1.995(8)	&  2.006(7)  \\
	&  O8	&  2.319(3)	&  2.319(4)	&  2.319(4)	&  2.307(6)	&  2.302(6)  \\
                                                                         \\
Cu2	&  O6	&  1.914(3) &  1.916(4)	&  1.918(4)	&  1.910(6)	&  1.913(6)  \\
	&  O4	&  1.939(3)	&  1.933(4)	&  1.932(4)	&  1.930(7)	&  1.940(6)  \\
	&  O2	&  1.979(4)	&  1.964(5)	&  1.970(4)	&  1.989(7)	&  1.977(7)  \\
	&  O3	&  1.995(4)	&  2.004(5)	&  2.000(6)	&  1.976(8)	&  1.983(8)  \\
	&  O8	&  2.406(3)	&  2.397(4)	&  2.401(4)	&  2.391(6)	&  2.385(6)  \\
                                                                         \\
Cu3	&  O1	&  1.966(3)	&  1.973(4)	&  1.964(4)	&  1.960(7)	&  1.970(6)  \\
	&  O3	&  1.970(3)	&  1.961(4)	&  1.966(5)	&  1.970(7)	&  1.961(7)  \\
	&  O2	&  1.992(3)	&  1.988(4)	&  1.980(4)	&  1.978(7)	&  1.984(6)  \\
	&  O7	&  1.997(3)	&  1.991(4)	&  1.999(4)	&  1.997(7)	&  1.990(7)  \\
	&  O8	&  2.368(4)	&  2.368(5)	&  2.375(5)	&  2.376(7)	&  2.370(6)  \\
                                                                         \\
Cu4	&  O7	&  1.974(4)	&  1.969(4)	&  1.967(4)	&  1.982(7)	&  1.980(7)  \\
	&  O2	&  1.993(4)	&  2.003(4)	&  1.994(4)	&  1.987(7)	&  1.984(6)  \\
	&  O3	&  2.006(4)	&  1.996(4)	&  2.011(5)	&  2.000(7)	&  1.996(7)  \\
	&  O1	&  2.031(3)	&  2.035(4)	&  2.035(5)	&  2.024(7)	&  2.025(6)  \\
	&  O4	&  2.351(4)	&  2.353(4)	&  2.348(5)	&  2.347(8)	&  2.344(7)  \\
	&  O6	&  2.398(4)	&  2.389(5)	&  2.398(5)	&  2.397(7)	&  2.372(7)  \\
                                                                         \\
S1	&  O11	&  1.470(4)	&  1.475(4)	&  1.480(5)	&  1.506(7)	&  1.488(63) \\
	&  O12	&  1.480(5)	&  1.482(5)	&  1.473(5)	&  1.482(8)	&  1.491(7)  \\
	&  O10	&  1.481(5)	&  1.484(5)	&  1.487(6)	&  1.485(8)	&  1.493(7)  \\
	&  O8	&  1.498(4)	&  1.499(5)	&  1.491(5)	&  1.500(7)	&  1.501(6)  \\
\end{tabular}
\end{ruledtabular}
\end{table}

\begin{table}[h]
\begin{ruledtabular}
\caption{\label{J_suppl} 
The total nearest-neighbor $J_{ij}$ (K) calculated with the HSE06 hybrid functional as implemented in \texttt{VASP5.2} in comparison with the results from LSDA+$U$ calculations using the \texttt{fplo9.07-41} code, $U_d=8.5\pm1$\,eV and $J_d=1$\,eV. d(Cu--Cu) and Cu--O--Cu denote Cu--Cu distances (\r{A}) and Cu--O--Cu angles (deg), respectively.}
\begin{tabular}{c c c c c}
          &  d(Cu--Cu)  &   Cu--O--Cu    & HSE06      & LSDA+$U$        \\ \hline
$J_s$     & 2.983       &	101.05	     &  $-116$    &  $-74\pm10$     \\
$J_1'$    & 3.011       &	97.81/98.64	 & 	  $21$    &    $9\pm15$     \\
$J_1$     & 3.020       &	99.10/99.49	 &    $58$    &   $38\pm20$     \\
$J_s'$    & 3.049       &	104.706	     &   $-65$    &  $-23\pm3$      \\
$J_a$     & 3.139       &	103.55       &   $-52$    &  $-19\pm1$      \\
$J_b$     & 3.144       &	105.08       &   $-23$    &    $5\pm5$      \\
$J_c$     & 3.163       &	106.64       &   $-43$    &   $-6\pm2$      \\
$J_d$     & 3.168       &	105.23       &   $-16$    &    $8\pm5$      \\
$J_e$     & 3.186       &	106.65       & 	 $-33$    &    $0\pm4$      \\
$J_f$     & 3.190       &	106.61       &   $-37$    &  $-12\pm2$      \\
$J_g$     & 3.219       &	106.10       &   $-37$    &  $-13\pm1$      \\
$J_h$     & 3.229       &	109.23       & 	   $5$    &   $31\pm7$      \\
\end{tabular}
\end{ruledtabular}
\end{table}

\clearpage

\begin{figure}[h]
\includegraphics[width=15cm]{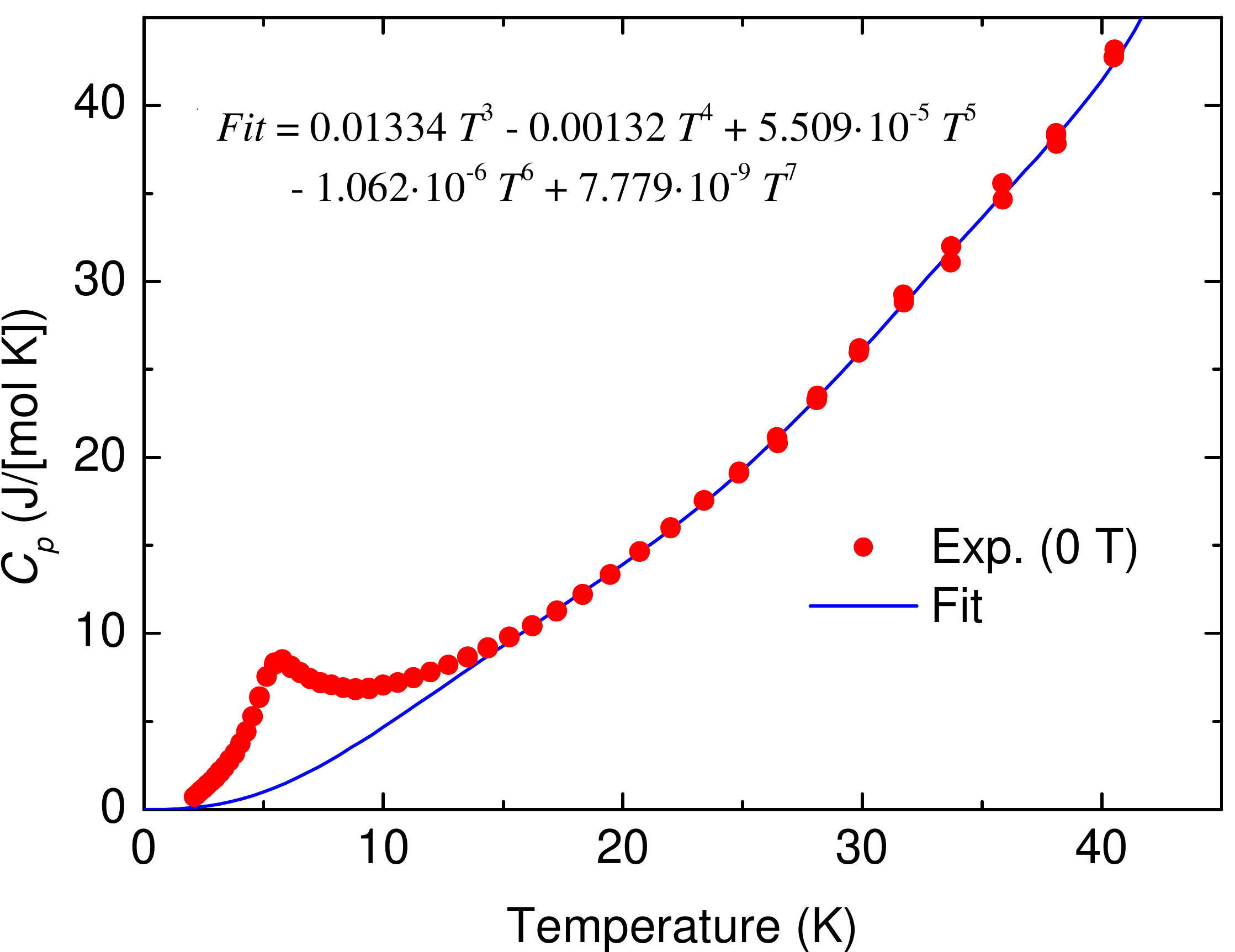}
\caption{\label{cp_bg}
(Color online) Specific heat data $C_p(T)$of Langite collected in zero magnetic field. The blue line shows the lattice background $C_{lat}(T)$ which we obtained by fitting a polynomial of the form $C_{lat}(T)=\sum \limits_{n=3}^{n=7} c_nT^n$ to the experimental data in the temperature regime 20--40\,K.  $C_p(T)-C_{lat}(T)$ yields the magnetic contribution $C_{mag}$ to the specific heat.}
\end{figure}

\begin{figure}[h]
\includegraphics[width=15cm]{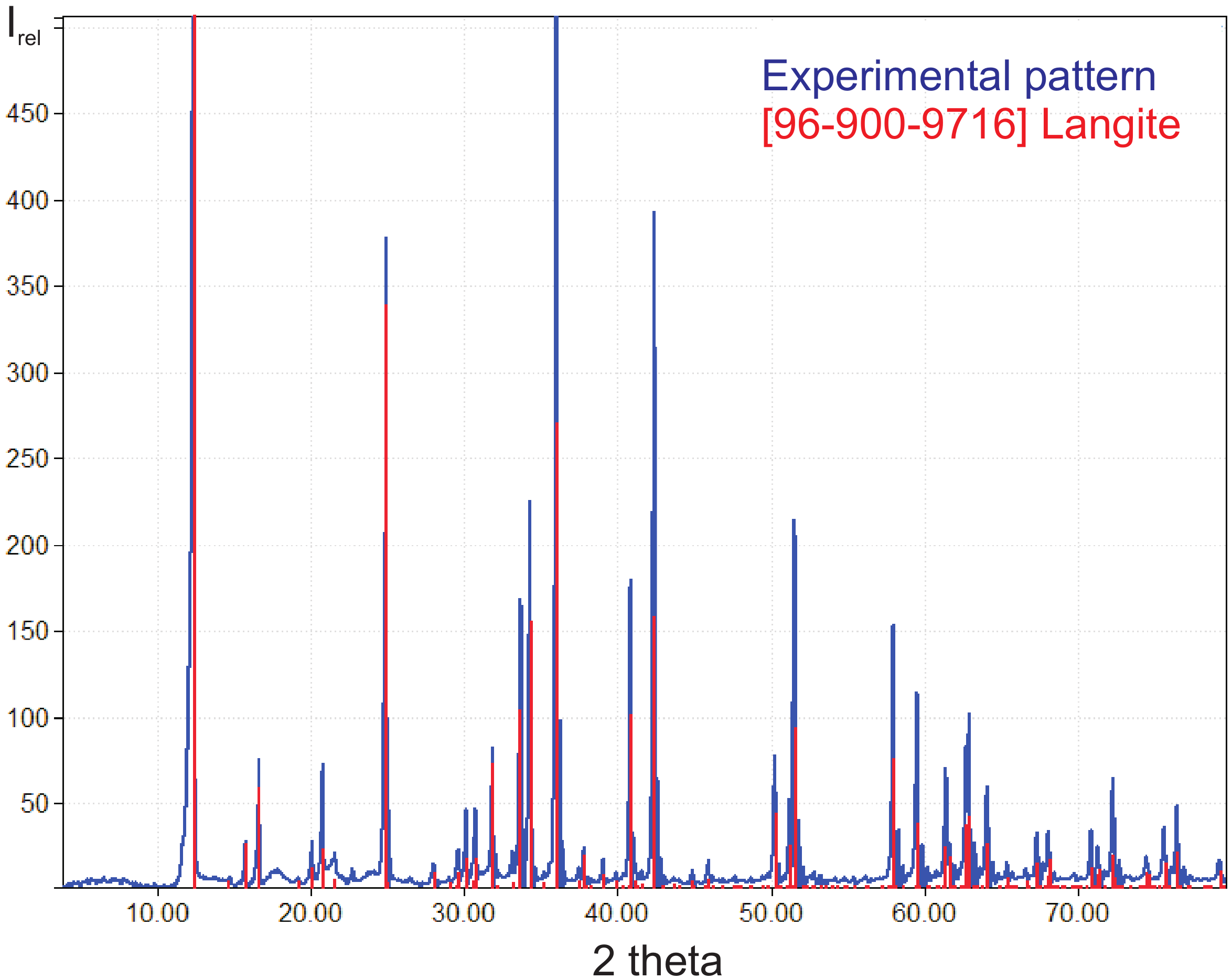}
\caption{\label{xrd}
(Color online) Room temperature powder X-ray diffraction pattern (Huber G670 Guinier camera, CuK$_{\alpha\,1}$ radiation, ImagePlate detector, $2\theta\,=\,3-100^{\circ}$ angle range) of the Langite sample.}
\end{figure}

\begin{figure}[h]
\includegraphics[width=16cm]{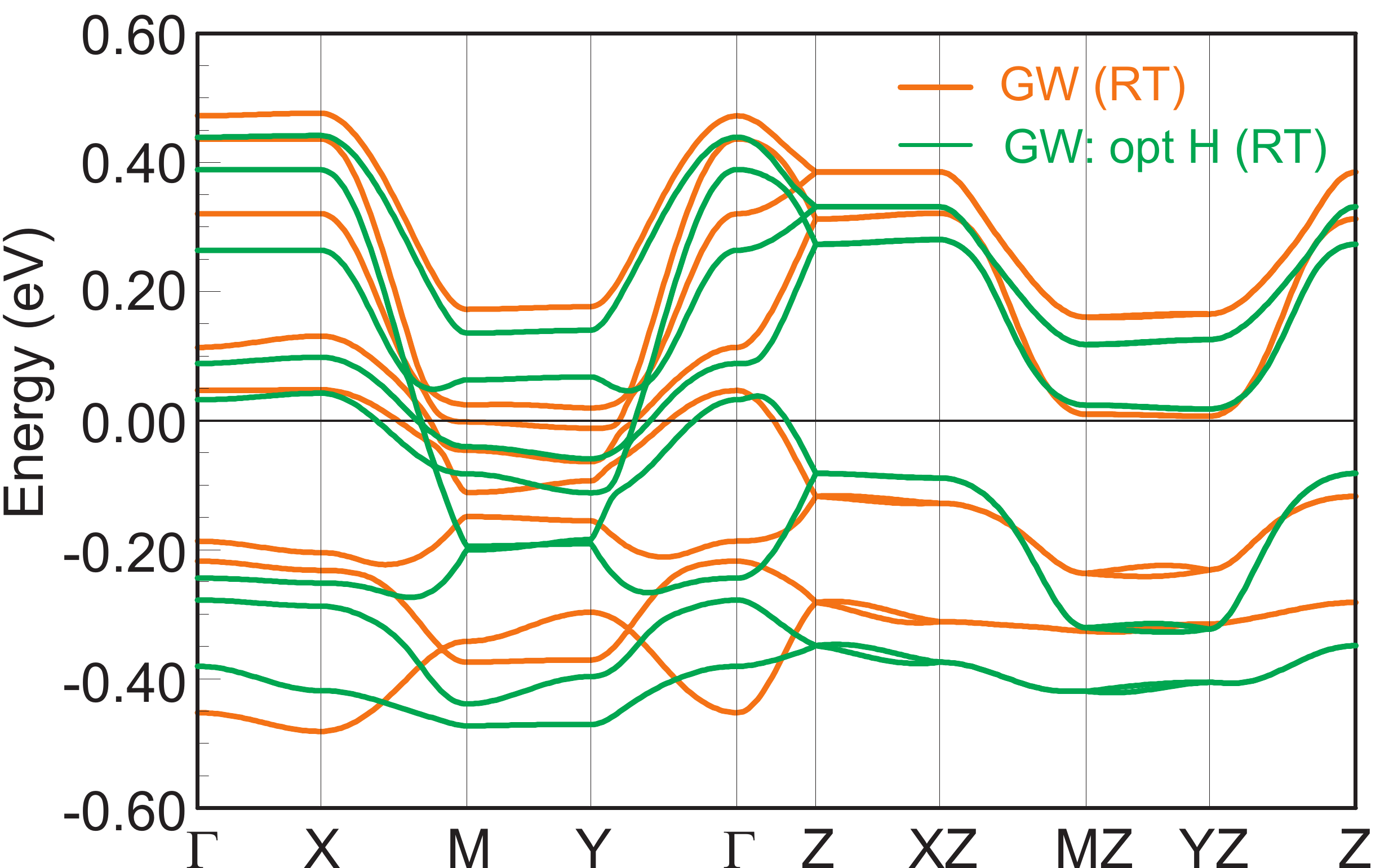}
\caption{\label{bandcomp3}
(Color online) The LDA band structure (orange) computed for the crystal structure provided by Gentsch and Weber in 1984 (GW). Their data was collected at room temperature (RT). The bands computed for GGA-optimized H-positions (opt H), with all other crystallographic data being fixed to those provided by GW, are shown in green. The optimization stabilizes bands up to about 50\,meV and significantly changes the band dispersion which is particularly pronounced in the energy region between -0.5 and -0.2\,eV.}
\end{figure}

\begin{figure}[h]
\includegraphics[width=16cm]{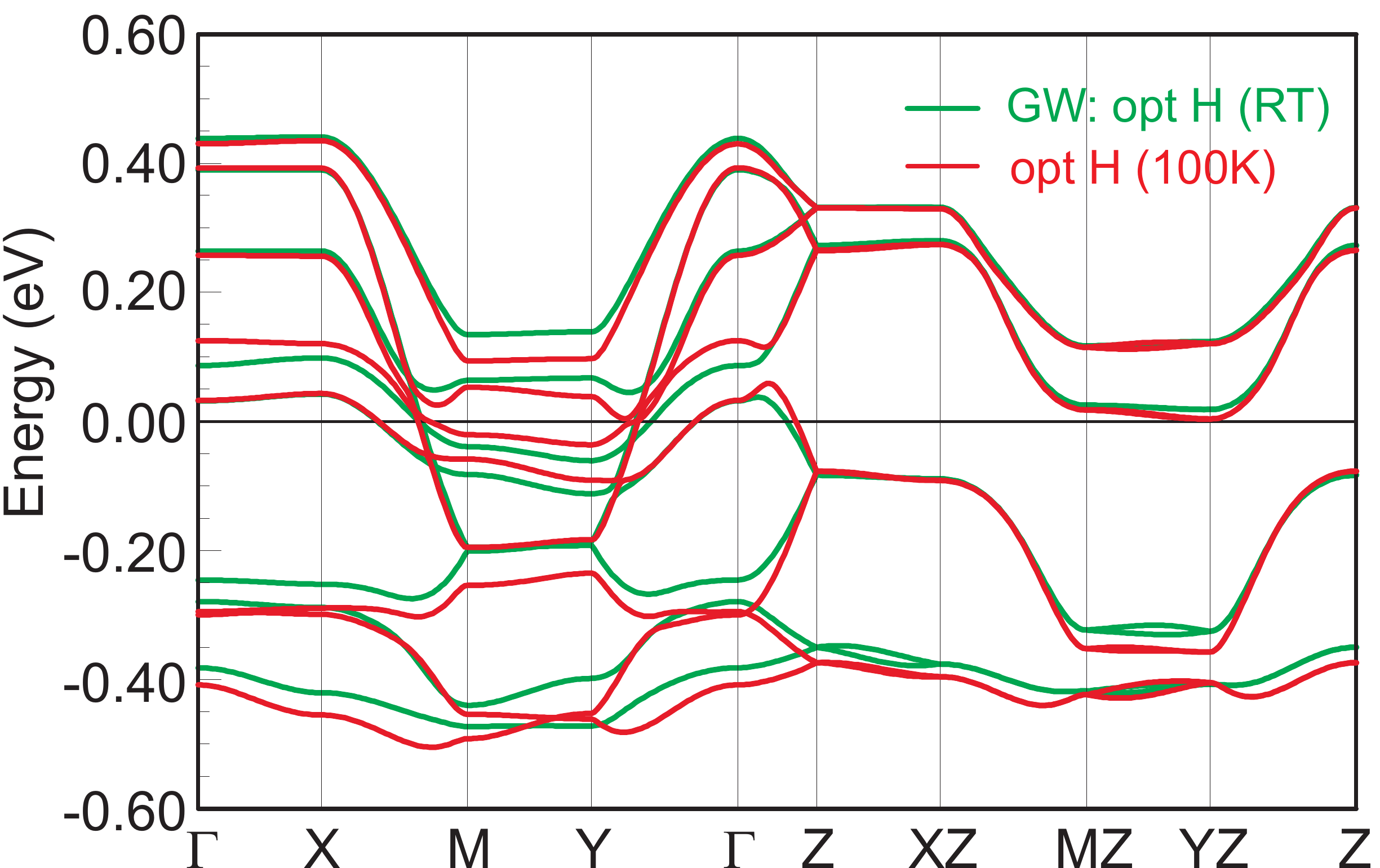}
\caption{\label{bandcomp1}
(Color online) The LDA band structure (green) computed for the crystal structure provided by Gentsch and Weber in 1984 (GW) with optimized H-positions (opt H). The red bands are obtained from our crystallographic data with optimized H-positions (opt H). In both cases one H is located at the SO$_4$ group. The differences probably arise from the different temperatures. While GW collected their data at room temperature (RT) ours was measured at 100\,K.}
\end{figure}

\begin{figure}[h]
\includegraphics[width=16cm]{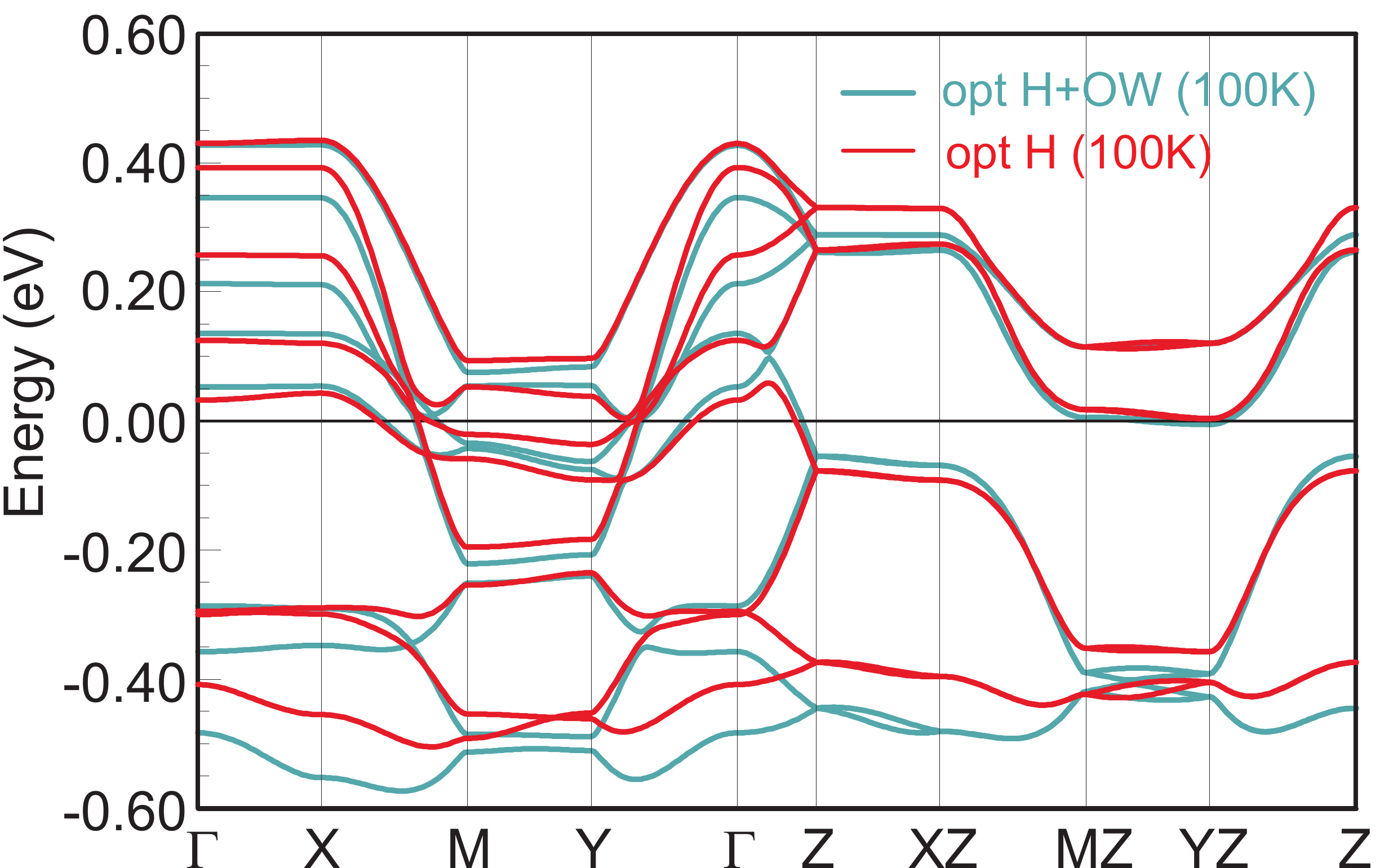}
\caption{\label{bandcomp2}
(Color online) The LDA band structure (red) computed for our crystal structure with optimized H-positions (opt H). The blue bands correspond to the structure where H and the oxygen atoms of the water molecules (OW) are optimized. In the first case one H is located at the SO$_4$ group. While in the latter case (opt H+OW) H from the SO$_4$ group has moved to the Cu-O layers. Particularly in the energy regime between -0.6 to -0.2\,eV band shifts up to 100\,meV are visible as well as considerable changes in the band dispersion. The structure with no H on SO$_4$ is considerably more stable.}
\end{figure}

\begin{figure}[h]
\includegraphics[width=12cm]{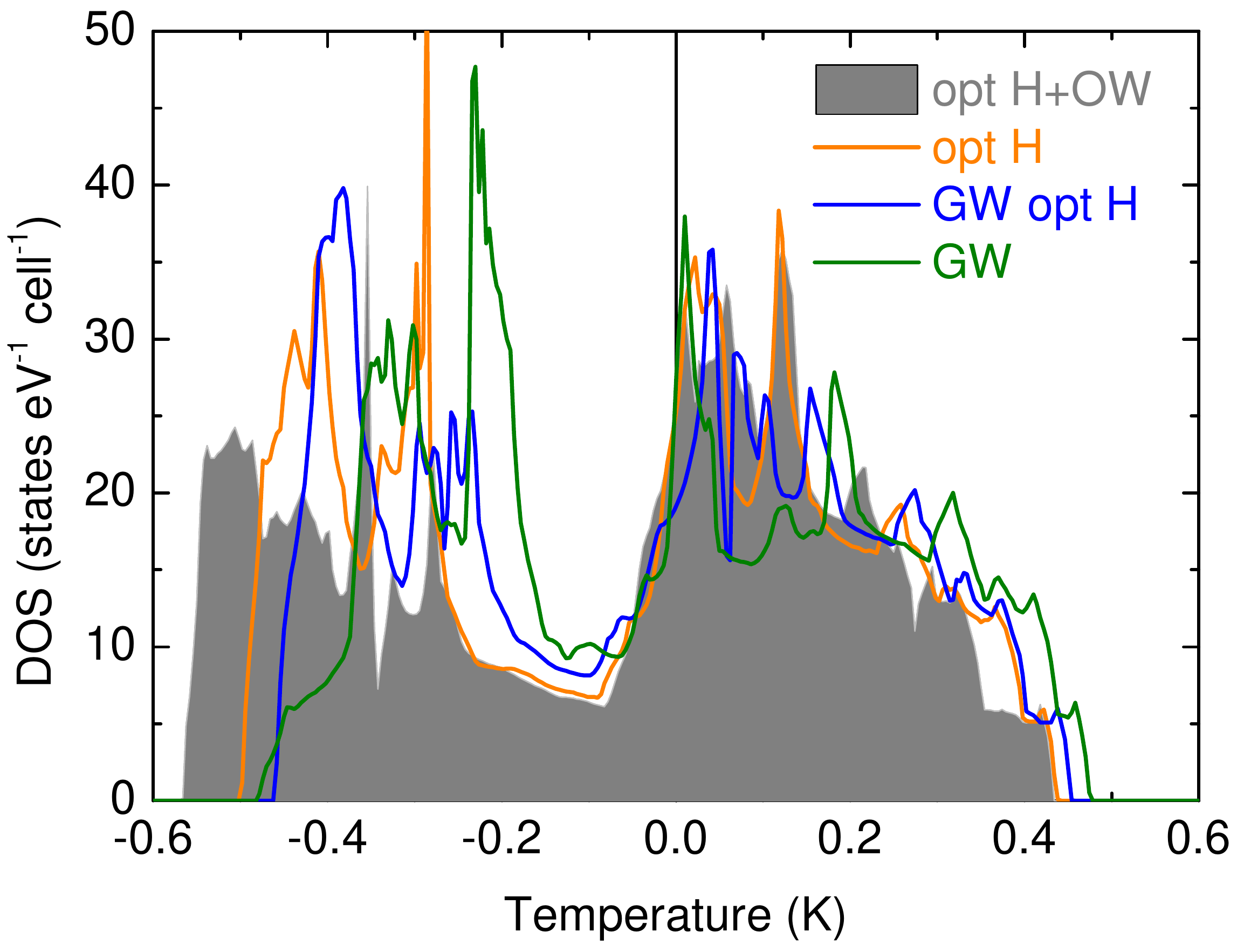}
\caption{\label{doscomp}
(Color online) A comparison of the density of states around the Fermi level calculated within the LDA for different crystal structures. "GW": the structure refined by Gentsch and Weber in 1984 at room temperature. "GW opt H": The GW crystal structure with GGA-optimized H-positions. "opt H" denotes our crystallographic data collected at 100\,K with optimize H-positions and "opt H+OW" that with optimized H and OW positions, where OW is the oxygen of the water molecules.}
\end{figure}

\begin{figure}[h]
\includegraphics[width=12cm]{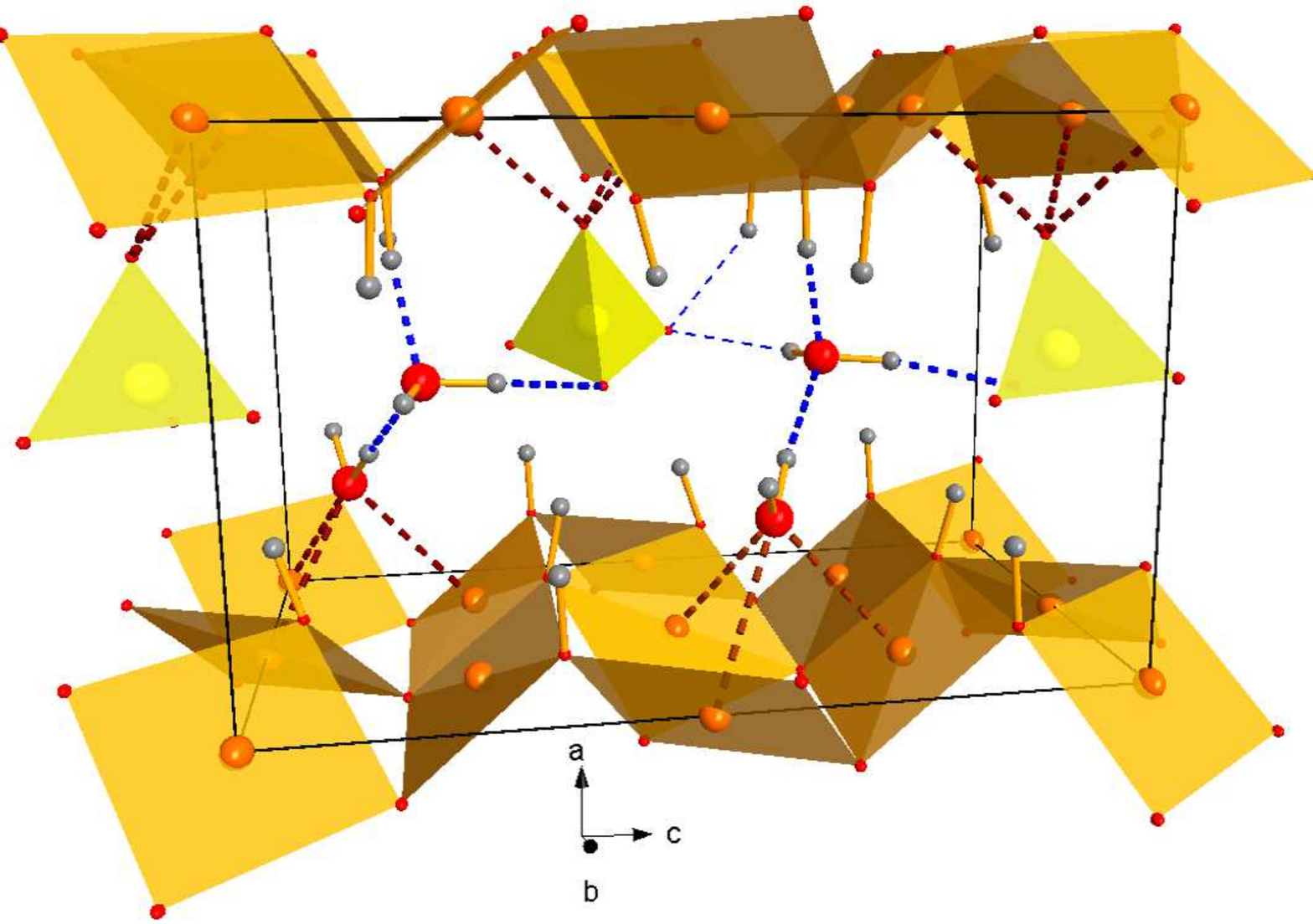}
\caption{\label{struct_ang}
(Color online) The bonding of the H$_2$O molecules and the SO$_4$ groups. Thick and thin dashed blue lines denote H-bonds shorter and longer than 0.80\,\r{A}. The dashed brown lines indicate the weak bonds between O atoms belonging to H$_2$O or SO$_4$ and Cu atoms.}
\end{figure}

\begin{figure}[h]
\includegraphics[width=15cm]{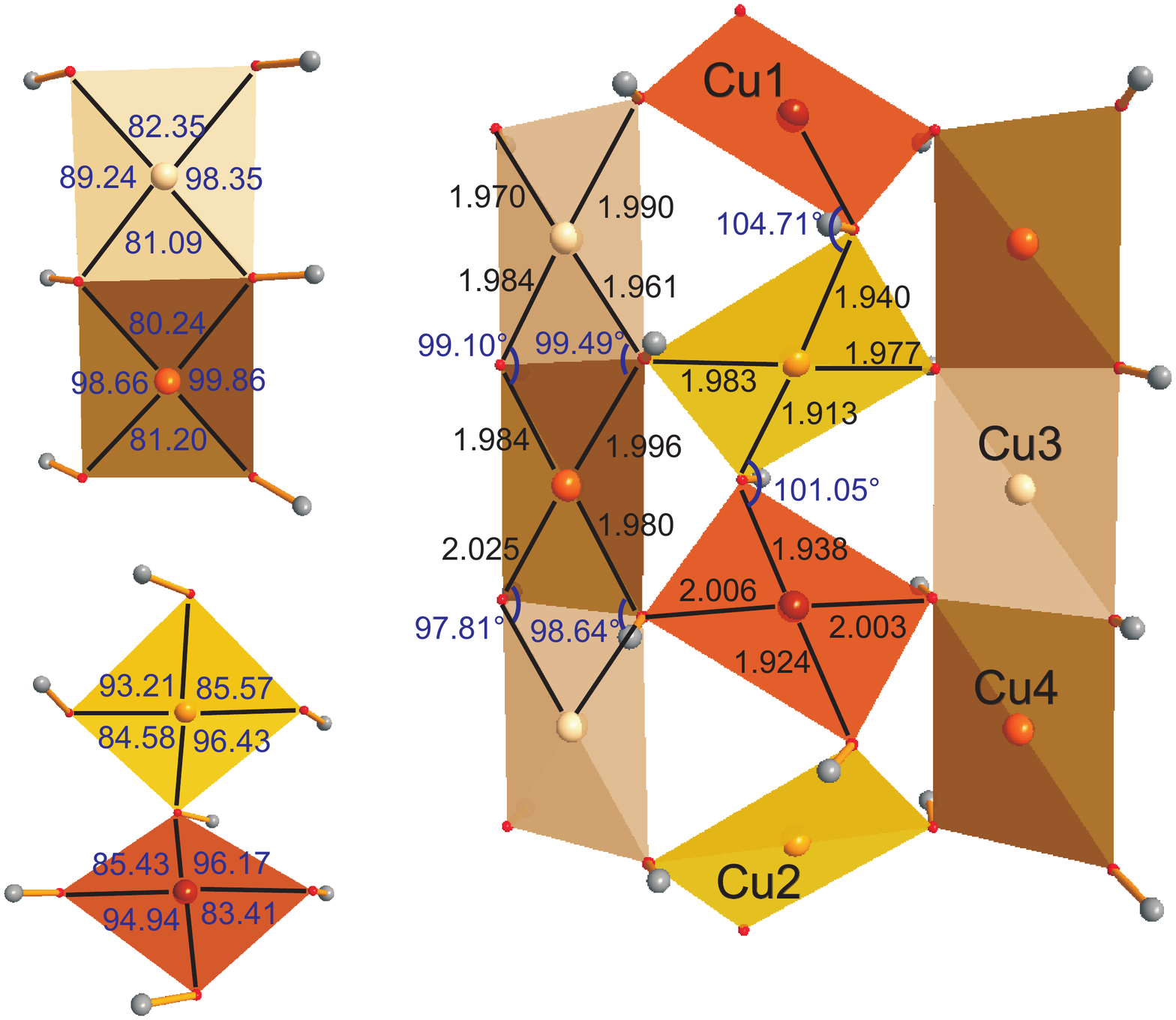}
\caption{\label{struct_ang}
(Color online) The right figure shows the structural layer in langite with the four different types of CuO$_4$ plaquettes. Black lines indicate Cu--O bonds. The black numbers with three digits denote the corresponding Cu--O bond lengths (in \r{A}). Blue numbers with two digits give the Cu--O--Cu bond angles in deg. To the left, fragments of type-A and type-B chains are shown, where the numbers give the O--Cu--O angles (in deg) on the different plaquettes.}
\end{figure}

\clearpage

\begin{figure} [h]
\includegraphics[width=17cm]{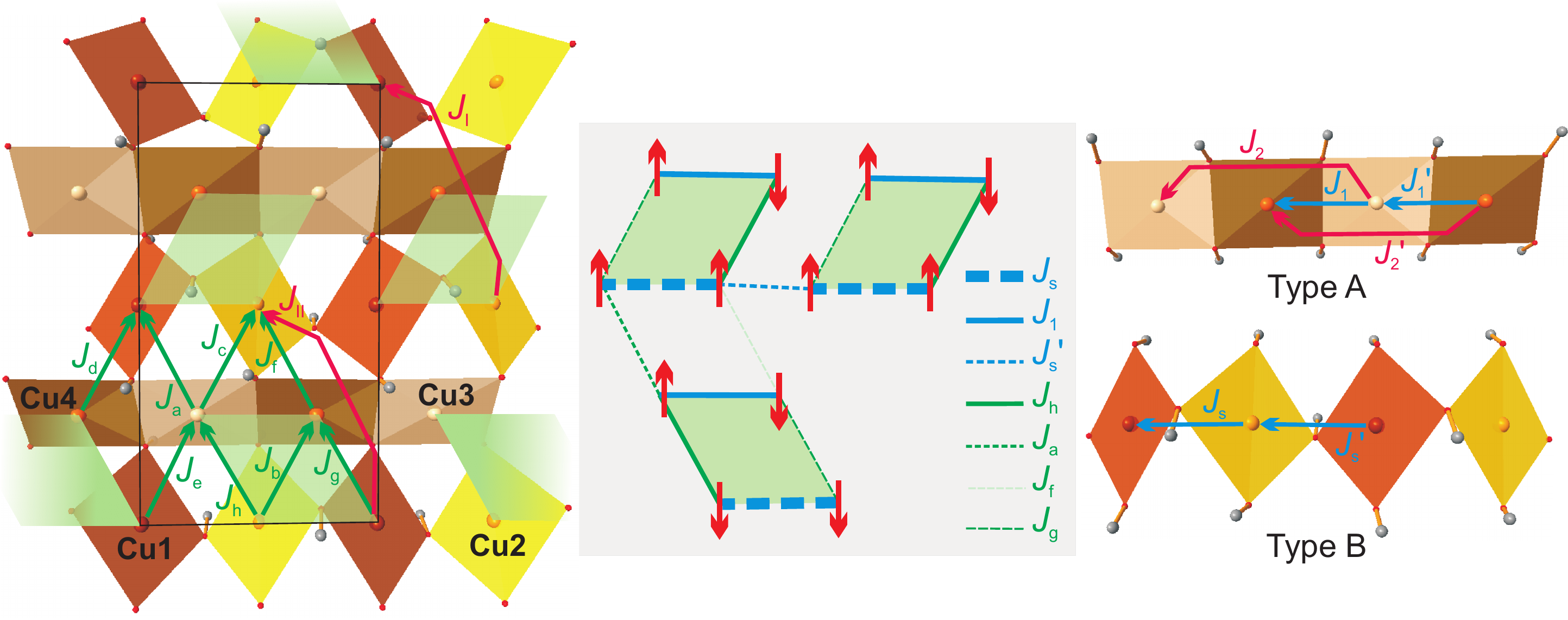}
\caption{\label{tetra}
(Color online) The left panel shows the intralayer exchange pathways. Transparent green areas represent the positions of the $S=1$ tetramers. These tetramers and the belonging nearest-neighbor $J_{ij}$ are shown in the central graphic where solid and dashed lines indicate AFM and FM interactions, respectively. $J_f$ is depicted in light green color since this coupling is frustrated. The weak couplings $J_b$ and $J_d$ are not displayed. Red arrows indicate the alignment of the $S_z=1/2$ spins for the isotropic model. The right panel shows the two chain types and the respective intrachain exchange pathways.}
\end{figure}

\begin{minipage}{17cm}
\textbf{Description of the tetramer model}

Considering first only the nearest-neighbor couplings (see Fig.~\ref{tetra}), an almost unfrustrated microscopic magnetic model for langite is obtained. Only the weak couplings $J_b$ and $J_f$ cannot be satisfied. $J_s$ and $J_s'$ are both FM with $J_s'$ being much smaller than $J_s$ rendering the B-type chains alternating ferromagnetic in this simple model. In the edge-sharing type A chains $J_1$ is AFM and about four times larger than the weak $J_1'$. Thus, the A-type chains can be reduced to AFM Cu3-Cu4 dimers. $J_h$, reigning between Cu2 and Cu3, couples these dimers to the ferromagnetic B-type chains resulting in $S=1$ tetramers (central panel in Fig.S3). Cu1 and Cu3 are coupled ferromagnetically by $J_a$ resulting in an antiparallel alignment of the tetramers along $c$. $J_s'$, in turn, is responsible for a parallel alignment along the chain direction $b$. Now, next-nearest neighbor exchanges are taken into account: $J_I$ and $J_{II}$ can be satisfied within the tetramer model but $J_2'$ and $J_2$ are frustrated.
The model of coupled $S=1$ tetramers allows the following phenomenological interpretation of $\chi(T)$ and M(H). The bend in M(H), starting at about 10\,T, arises from a parallel alignment of the $S=1$ tetramers in the magnetic field. In this case, 3/4 of the spins are oriented parallel to the magnetic field resulting in a magnetization of 0.5\,$\mu_B$/Cu in agreement with the measurements. The effective AFM coupling between the tetramers $J_{it}$ can roughly be estimated by an AFM $S=1$ square-lattice spin model yielding $8J_{it}k_B/(g\mu_B) \approx 10$\,T and accordingly $J_{it} \approx 1.8$\,K. This effective coupling is considerably smaller than our calculated values, e.g. for $J_I$ and $J_{II}$, however, the seeming discrepancy can be explained by frustration resulting in the very weak effective intertetramer coupling. A similarly small value for $J_{it}$ can be obtained from $\chi(T)$. The maximum of the magnetic susceptibility for an AFM $S=1$ square-lattice is at about $2.18T/J_{it}$. According to the experimental $\chi(T)_{\text{max}}=7.5$\,K, we, thus, obtain $J_{it}=3.4$\,K. With respect to our crude estimate based on an AFM $S=1$ square-lattice the agreement between the $J_{it}$ extracted from $\chi(T)$ and M(H) is reasonable and, therefore, justifies the $S=1$ tetramer model and also clearly shows the crucial role of frustration in langite.  

We employ again the tetramer model and assume an intratetramer coupling of about 20--40\,K. Now, a saturation field may roughly be estimated using a two-dimensional system with four effective couplings on each Cu-site yielding $4Jk_B/(g\mu_B) \approx 60-120$\,T. The bend at about 4\,T cannot easily be ascribed to certain mechanisms within the phenomenological model but may arise from secondary effects. 

The model of coupled tetramers cannot provide a better description of the magnetic properties than the model of coupled magnetic sublattices which we present in the main text. It also neglects the strong NNN $J_2$ and $J_2'$ couplings and appears to be counterintuitive since chain features are completely lost. Such feature, however, were observed in neutron experiments on related compounds like brochantite and antlerite (see main text). Therefore, it appears natural to stick to a model based on magnetic chains as long as it is not proved inappropriate.
\end{minipage}

\end{widetext}

\end{document}